\documentclass[useAMS,usenatbib]{mn2e}

\usepackage{graphicx}
\usepackage{amssymb}
\usepackage{epsfig}
\usepackage{amssymb}
\usepackage{commath}
\usepackage{adjustbox}

\title[]{The Ophiuchus DIsc Survey Employing ALMA (ODISEA) - I : project description and continuum images at 28 au resolution}
\author[Lucas A. Cieza  et al.]
{
\Large 
\vspace{0.15cm}
Lucas A. Cieza$^{1,2}$, 
Dary Ru\'iz-Rodr\'iguez$^{3}$,
Antonio Hales$^{4,5}$, 
Simon Casassus$^{2,6}$, 
Sebastian P\'erez$^{2,6}$,
\\
\vspace{0.15cm}
\Large 
Camilo Gonzalez-Ruilova$^{2,6}$,
Hector C\'anovas$^{7}$,  
Jonathan P. Williams$^{8}$, 
Alice Zurlo$^{1,2}$,   
Megan Ansdell$^{9}$, 
\\
\vspace{0.15cm}
\Large 
Henning Avenhaus$^{10}$,
Amelia Bayo$^{11,12}$,  
Gesa H. -M. Bertrang$^{2,6,13}$,
Valentin Christiaens$^{2, 6}$,
William Dent$^{4}$,  
\\
\vspace{0.15cm}
\Large 
Gabriel Ferrero$^{14}$,
Roberto Gamen$^{14}$,
Johan Olofsson$^{11,12}$,
Santiago Orcajo$^{14}$,
Karla Pe\~na Ram\'irez$^{15}$,
\\
\vspace{0.15cm}
\Large 
David Principe$^{16}$, 
Matthias R. Schreiber$^{11,12}$,
Gerrit van der Plas$^{17}$\\
$^{1}$Facultad de Ingenier\'ia y Ciencias, N\'ucleo de Astronom\'ia, Universidad Diego Portales, Av. Ejercito 441. Santiago, Chile \\
$^{2}$Millennium Nucleus ``Protoplanetary Discs in ALMA Early Science", Av. Ejercito 441. Santiago, Chile \\
$^{3}$Chester F. Carlson Center for Imaging Science, Rochester Institute of Technology, Rochester, NY 14623-5603, USA \\
$^{4}$Joint ALMA Observatory, Alonso de Cordova 3107, Vitacura 763-0355, Santiago, Chile \\ 
$^{5}$National Radio Astronomy Observatory, 520 Edgemont Road, Charlottesville, Virginia, 22903-2475, USA \\
$^{6}$Departamento de Astronom\'ia, Universidad de Chile, Casilla 36-D Santiago, Chile \\
$^{7}$European Space Astronomy Centre (ESA), Camino Bajo del Castillo s/n, 28692, Villanueva de la Ca\~nada, Madrid, Spain \\
$^{8}$Institute for Astronomy, University of Hawaii at Manoa, Honolulu, HI 96822, USA \\
$^{9}$Department of Astronomy, University of California at Berkeley, Berkeley, CA 94720-3411 \\
$^{10}$ETH   Zurich,  Institute for Particle Physics and Astrophysics,   Wolfgang-Pauli-Strasse 27, CH-8093, Zurich, Switzerland \\
$^{11}$Facultad de Ciencias, Instituto de F\'isica y Astronom\'ia, Universidad de Valpara\'iso, Av. Gran Breta–a 1111, 5030 Casilla, Valpara\'so, Chile \\
$^{12}$Millennium Nucleus for Planet Formation, Universidad de Valpara\'iso, Av. Gran Breta\~na 1111, Valpara\'iso, Chile \\
$^{13}$Max Planck Institute for Astronomy, K\"onigstuhl 17, 69117 Heidelberg, Germany \\
$^{14}$Instituto de Astrof\'isica de La Plata y Facultad de Ciencias Astron\'omicas y Geof\'isicas, UNLAP, Paseo del Bosque s/n, La Plata, Argentina \\
$^{15}$Centro de Astronom\'ia (CITEVA), Universidad de Antofagasta, Av. Angamos 601,  Antofagasta, Chile \\
$^{16}$Massachusetts Institute of Technology, Kavli Institute for Astrophysics, Cambridge, MA, USA\\
$^{17}$Univ. Grenoble Alpes, CNRS, IPAG, F-38000 Grenoble, France \\
}
\begin{document}
\pagerange{\pageref{firstpage}--\pageref{lastpage}} \pubyear{2017}

\maketitle

\label{firstpage}

\begin{abstract}
We introduce the Ophiuchus DIsc Survey Employing ALMA (ODISEA),  a project aiming to study the entire population of \emph{Spitzer}-selected protoplanetary discs in the Ophiuchus Molecular Cloud ($\sim$300 objects) from both millimeter continuum and CO isotopologues data.   
Here we present 1.3 mm/230 GHz continuum images of 147 targets at 0.2$''$ (28 au) resolution and a typical rms of 0.15 mJy. We detect a total of 133 discs, including the individual components of 11 binary systems and 1 triple system. Fifty-three of these discs are spatially resolved. We find clear substructures (inner cavities, rings, gaps, and/or spiral arms) in 8 of the sources and hints of such structures in another 4 discs. 
We construct the disc luminosity function for our targets and perform comparisons to other regions. 
A simple conversion between flux and dust mass (adopting standard assumptions) indicates that all discs detected at 1.3 mm  are massive enough to form one or more rocky planets.  
In contrast, only $\sim$50 discs ($\sim$1/3 of the sample) have enough mass in the form of dust to form the canonical 10 M$_{\oplus}$ core needed to trigger runaway gas accretion and the formation of gas giant planets, although the total mass of solids already incorporated into bodies  larger than cm scales is mostly unconstrained. 
The distribution in continuum disc sizes in our sample is heavily weighted towards compact discs:  most detected discs have radii  $<$ 15 au, while only 23 discs ($\sim$15$\%$ of the targets) have 
radii $>$ 30 au. 
\end{abstract}


\section{Introduction}\label{introduction}

The diversity and high incidence of extrasolar planets in the field (Cassan et al. 2012; Howard, 2013; Burke et al. 2015; Shvartzvald et al. 2016)  demonstrate that  planet formation processes are efficient  and imply that most of the circumstellar discs  seen in nearby molecular clouds should form planetary systems. 
Studying the structure and evolution of complete populations of  protoplanetary discs in these clouds is  thus  important to place constraints
on the conditions, timescales, and mechanisms associated with planet formation.   
Early-science observations of individual protoplanetary discs  with the Atacama Large Millimeter/submillimeter Array (ALMA) have produced transformational results (Van Der Marel. et al. 2013;  Casassus et al. 2013; ALMA Partnership et al. 2015; Perez et al. 2016; Cieza et al. 2016; Andrews et al. 2016). 
However,  detailed submillimeter studies tend to target bright sources (F$_{mm}$  $\gtrsim$ 50 mJy) and are therefore very biased towards massive discs around relatively massive stars. 
Also, many imaging studies have focused on bright ``transition objects" (e.g., massive discs with inner holes and gaps tens of au wide). 
While very important, such objects are \emph{not} representative of the typical planet-forming disc in a molecular cloud; they only represent $\lesssim$ 10$\%$ of the young disc population (Cieza et al. 2012a).
The disc population is very diverse, which might reflect a wide range of initial conditions (Bate et al. 2018) and evolutionary paths (Cieza et al. 2007; Currie $\&$ Kenyon 2009).
Transition discs with large cavities could be related to the formation of multiple giant planets (Owen et al. 2016), which are rare according to extrasolar planet studies.  
On the other hand, planetary systems with low-mass planets are much more common in the Galaxy.
In particular, Gaidos et al. (2016) estimate that M-dwarfs,  the most common type of  star in the Milky Way, host an average of 2.2 $\pm$ 0.3 planets with radii of 1-4 R$_{\oplus}$  and orbital periods less than 180 days. 
Such planets could in principle form in discs that are only a few au in radius and contain just a few Earth masses of dust. 

Previous infrared  surveys with \emph{Spitzer}, tracing mostly optically thick emission,  have shown that the \emph{presence} of a disc is a strong function of stellar age and that protoplanetary discs have a mean lifetime of  $\sim$3 Myr (Williams $\&$ Cieza, 2011).
ALMA's unprecedented  sensitivity provides, for the first time,  the opportunity to study complete samples of discs  at sub-arcsecond resolution in the (sub)millimeter regime, where discs become optically thin and resolved images  trace the  spatial distribution of mass. 
ALMA has already surveyed many of the nearby ($\lesssim$ 250 pc)  disc populations:  Ansdell et al. (2016; 2018) observed $\sim$90 discs in Lupus,  Barenfeld  et al. (2016) studied  106 objets in the Upper Scorpius OB Association,  and Pascucci et al. (2016) investigated 93 discs in the Chamaeleon I star-forming region. Similarly,  Ansdell et al. (2017) studied  92 objects in $\sigma$ Ori,  Cox et al.  (2017)  observed 49 systems in Ophiuchus, 
and Ruiz-Rodriguez et al. (2018) studied 136 discs in the IC~348 cluster. 
In general, these surveys observe gas tracers ($^{12}$CO, $^{13}$CO, and/or C$^{18}$O) and dust continuum. 
Since the dust continuum is easier to detect and study,  total disc masses are typically derived assuming a gas to dust mass ratio of 100. 
Under this assumption, surveys usually find a very wide range of discs masses ($<$ 1 to $\sim$100 M$_{\rm JUP}$) and a strong dependence of  disc mass on stellar mass. 
A clear overall decrease on disc mass with stellar age is also seen; however, some discs remain undetected at mm wavelengths at very young ages ($\lesssim$ 1 Myr) 
while a few massive discs are still seen in older regions ($\gtrsim$ 5 Myr).
Studies attempting to derive gas masses from CO isotopologues  often find gas-to-dust mass ratios significantly lower than 100,  but it remains to be stablished whether these results reflect  the depletion of total gas mass or just volatile carbon (Ansdell et al. 2016; Miotello et al. 2017).
\\

As part of the Ophiuchus DIsc Survey Employing ALMA (ODISEA) project, here we present Band-6  (230 GHz/1.3 mm) continuum observations of 147 discs in the Ophiuchus Molecular Cloud at a spatial resolution of 0.2$''$ (28 au). 
This is the largest (sub)millimeter disc study at this physical resolution to date and represents 50$\%$ of the full ODISEA sample (see Sec~\ref{sample}). 
Together with the Taurus Molecular cloud, Ophiuchus has been one of the best-studied regions in the (sub)millimeter regime in the pre-ALMA era and  it has played a central role in our understanding of protoplanetary disc populations.  Andrews $\&$ Williams (2007) presented single-dish submillimeter observations for 48 Ophiuchus sources (resolution = 14$''$ and  rms $\sim$5 mJy at 850 $\mu$m). They also collected 1.3 mm measurements for 99 additional targets  from the literature (rms $\sim$ 10 mJy), for a total sample of 147 objects (the same number as in this paper, but not necessarily all the same sources), resulting in 64 detected objects with estimated disc masses between 1 and 200 M$_{\rm JUP}$. They  estimate that the typical disc in their sample has 1$\%$ of the stellar mass. 
Cieza el al. (2010) presented (sub)millimeter photometry for 26 Ophiuchus ``transition discs", broadly defined there as objects with reduced levels of infrared excess with respect to the median found in T Tauri stars\footnote{Transition discs have been defined in several different ways in the literature,  see for example the \emph{diskionary} by Evans et al. 2009a; https://arxiv.org/pdf/0901.1691.pdf} with a similar disc mass sensitivity of $\sim$1  M$_{\rm JUP}$. 
They found that accreting transition objects tend to have discs that are bright in the (sub)millimeter, while non-accreting transition objects tend to have much fainter discs that are 
usually undetected at (sub)millimeter wavelengths.  
Andrews et al. (2009; 2010) used the Submillimeter Array (SMA) to image 17 of the brightest  Ophiuchus discs (75 $>$ mJy at 850 $\mu$m) at 0.3$''$ resolution and found that 4 of them had resolved inner cavities. 
More recently, Cox et al.  (2017) used ALMA to observe 49 Ophiuchus systems at 870 $\mu$m with a resolution of 0.2$''$ in dust continuum only, making it 
the largest survey of resolved protoplanetary discs in Ophiuchus before ODISEA. 
They find that binary systems tend to have smaller and lower mass discs and identified at least four objects with gaps and/or inner cavities in their discs. 

The ODISEA project aims to produce a complete demographic study of the discs in Ophiuchus to investigate disc evolution and the planet-formation potential of the entire cloud. 
This is the first of a series of papers that will also include  1) a study of disc properties as a function of the mass and age  of the host stars (Ruiz-Rodriguez et al. in preparation), 2) an investigation of the effects of (sub)stellar companions on disc properties (Zurlo et al. in preparation),  3) radiative transfer modeling of resolved sources (Perez et al. in preparation), and 4) a study of gas content in the discs based on $^{12}$CO, $^{13}$CO, and/or C$^{18}$O observations (Williams et al. in preparation). 
In  Section~2, we discuss the sample selection for the ODISEA project and our ALMA Cycle-4 observations.
In Section~3, we present our dust continuum images. We measure dust continuum fluxes and disc sizes for all resolved sources and provide deprojected radial profiles to search for discs showing sub-structures (inner cavities and gaps). 
In Section~4, we compare our results to those of previous surveys and  discuss their implications for disc evolution and planet formation. 
A summary of our main results and conclusions is presented in Section~5.

\begin{figure*}
\includegraphics[width=10cm, trim = 0mm 21mm 0mm 0mm, clip]{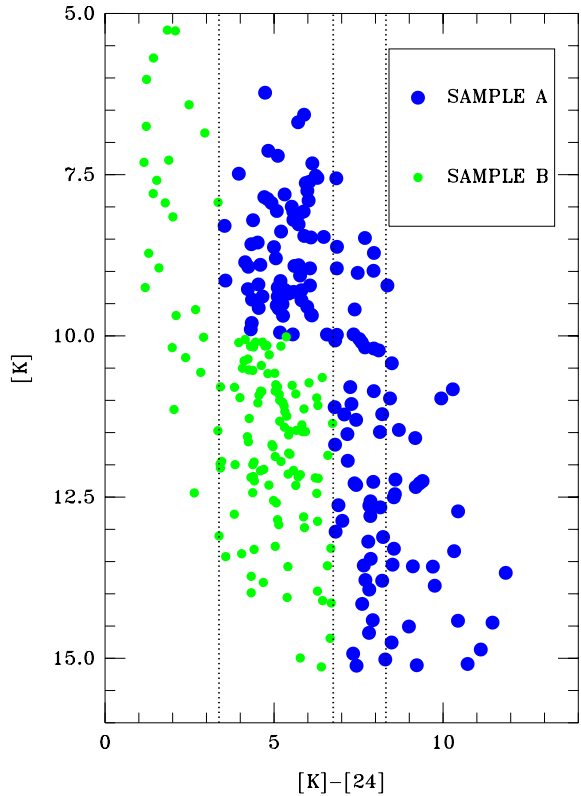}
\caption{ \small The  [K] vs [K]-[24] color-magnitude diagram of the full sample of \emph{Spitzer} Young Stellar Object Candidates in Ophiuchus identified by the \emph{Cores to Discs} Legacy Program. 
The vertical lines are the approximate boundaries between Class III, Class II, Flat Spectrum, and Class I objects (left to right). 
The \emph{Spitzer} objects were divided into two ODISEA samples. 
Sample ``A", presented in this paper, contains 147 objects, which are Class I and Flat Spectrum sources together with 
Class II objects brighter than 10 mag in K-band. 
}
\label{fig:sample}
\end{figure*}

\begin{table*}
\begin{center}
\caption[]{\em{ODISEA Cycle-4 SAMPLE}}
\begin{tabular}{@{}rlccccrrcr@{}}
\hline
 ID         &    C2D ID           &   RA       &    Dec &            RA      &  Dec         &    SpT  & Ref1 &    Separation & Ref2  \\
               &                     &  (J2000.0) & (J2000.0) &  (deg)        &     (deg)   &             &            &     ($''$) \\
\hline
  1  &  SSTc2d J162131.9-230140  & 16:21:31.920  & -23:01:40.25  & 245.382996  & -23.027847  &  ... &  ... &  ... &  ... \\
  2  &  SSTc2d J162138.7-225328  & 16:21:38.722  & -22:53:28.26  & 245.411346  & -22.891182  &  ... & ...  &  ... &  ... \\
  3  &  SSTc2d J162145.1-234232  & 16:21:45.127  & -23:42:31.63  & 245.438034  & -23.708786  &  ... &  ... &  ... &  ... \\
  4  &  SSTc2d J162148.5-234027  & 16:21:48.473  & -23:40:27.26  & 245.451965  & -23.674240  &  ... & ...  &  ... &  ... \\
  5  &  SSTc2d J162218.5-232148  & 16:22:18.521  & -23:21:48.12  & 245.577164  & -23.363367  &  K5 & 1  &  0.02 &   1 \\
  6  &  SSTc2d J162221.0-230402  & 16:22:20.990  & -23:04:02.35  & 245.587463  & -23.067320  &   ... & ...  &    &  ... \\
  7  &  SSTc2d J162225.0-232955  & 16:22:24.950  & -23:29:54.91  & 245.603958  & -23.498587  & ...  & ... &  ... &  ... \\
  8  &  SSTc2d J162245.4-243124  & 16:22:45.389  & -24:31:23.82  & 245.689117  & -24.523283  &    M3  &  1  &   0.54 &   2  \\
  9  &  SSTc2d J162305.4-230257  & 16:23:05.431  & -23:02:56.73  & 245.772629  & -23.049091  &   ...& ....  &  ... &  ... \\
 10  &  SSTc2d J162306.9-225737  & 16:23:06.859  & -22:57:36.61  & 245.778580  & -22.960171  &   ....&    ... &  ... &  ...  \\
\hline
\end{tabular}\label{table:sample}
\\
\noindent \scriptsize{Comments:  Only the first 10 lines are shown. The full table is available online. 
 References for spectral types are as follows:
1 =             Cieza et al. (2010);
2  =            Erickson et al. (2011);
3   =           Wilking et al. (2005);
4    =          Manara et al. (2015);
5     =         McClure et al. (2010);
6      =        Cieza et al. (2007);
7       =       Luhman et al. (1999).
References for spectral types are as follows:
1 = Ruiz-Rodriguez et al. 2016;
2 = Cieza et al. 2010;
3 =  Ratzka et al. 2005; 
4 = Loinard et al. 2008; 
5 = This work; 
6 = Kohn et al. 2016.
}
\end{center}
\end{table*}

\begin{figure*}
\includegraphics[width=18cm, trim = 0mm 0mm 0mm 0mm, clip]{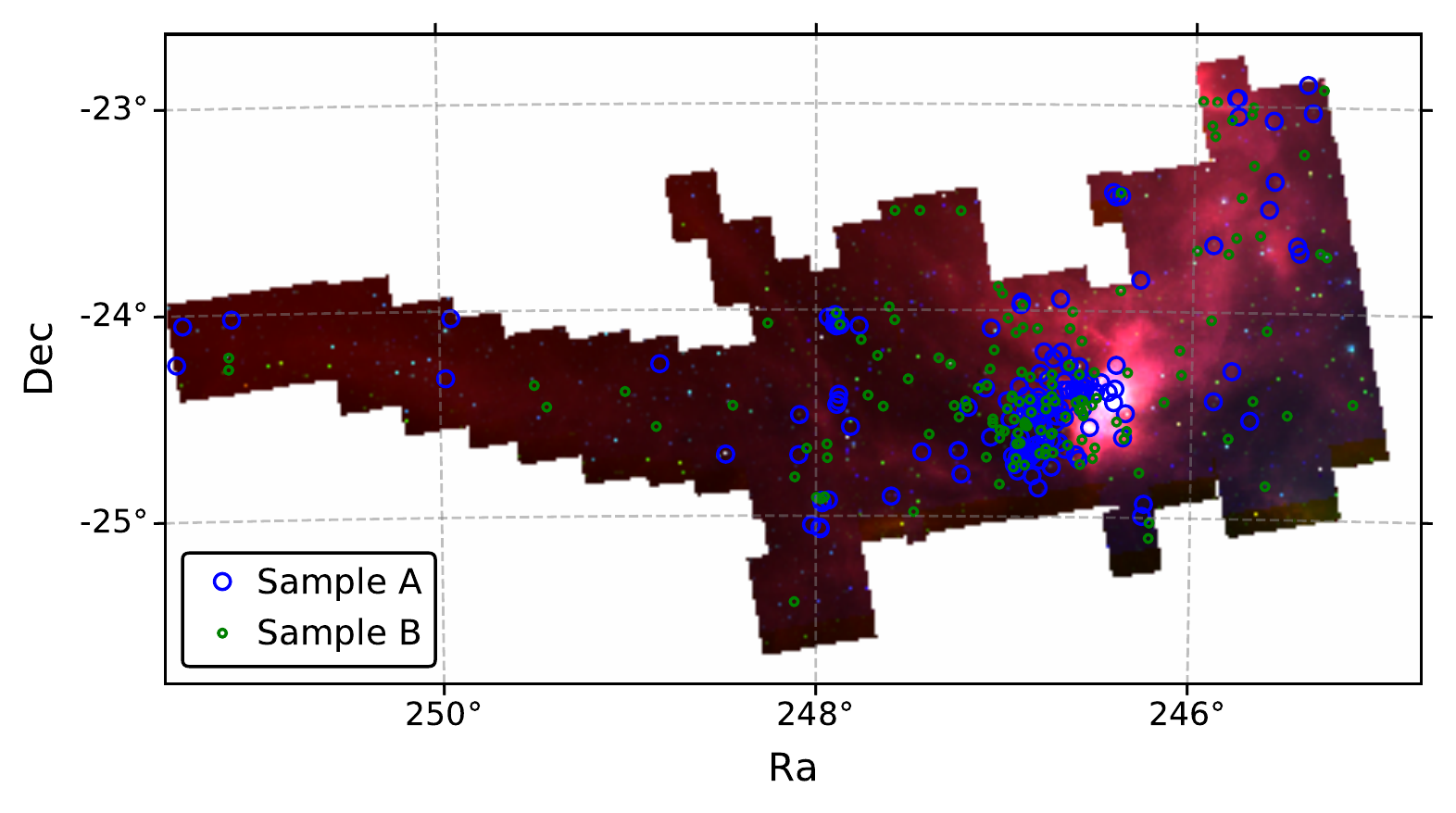}
\caption{ \small The spatial distribution of the ODISEA targets (both Samples A and B) shown on top of the \emph{Spitzer} map of the Ophiuchus molecular cloud from the ``Cores to Discs" Legacy Project.}
\label{fig:spitzermap}
\end{figure*}

\begin{table*}
\tiny
\begin{center}
\caption[]{\em{Photometry from the \emph{Spitzer} Cores to Discs Catalogue}}
\begin{tabular}{rrrrrrrrrrrrrrrrrrrr}
\hline \noalign {\smallskip}
ID & J &  eJ       &        H   & eH      &  K         & eK      & F$_{3.6}$ & eF$_{3.6}$ &   F$_{4.5}$  & eF$_{4.5}$  &  F$_{5.8}$  & eF$_{5.8}$ & F$_{8.0}$  & eF$_{8.0}$  & F$_{24}$ & eF$_{24}$ & F$_{70}$ & eF$_{70}$ \\ 
& \multicolumn{18}{c}{(mJy)}  \\
\hline \noalign {\smallskip}
1    &   36.10   &     0.86   &    80.60    &     1.93    &     103.00    &      2.00    &    101.00   &       5.03    &    85.00   &      4.23   &    73.90   &    3.51   &   79.40   &    3.86   &  119.0  &   11.0   &    175    &    25   \\
2    &    0.25   &     0.05   &     0.46    &     0.07    &       0.60    &      0.09    &      0.81   &       0.04    &     0.97   &      0.05   &     1.46   &    0.08   &    5.68   &    0.27   &   31.7   &    2.9   &      ...    &     ...   \\
3    &    0.50   &     0.05   &     1.33    &     0.08    &       2.47    &      0.12    &      5.50   &       0.28    &     9.11   &      0.45   &     7.26   &    0.36   &   14.30   &    0.70   &  201.0   &   18.6   &    704    &    87   \\
4    &    5.82   &     0.15   &    11.80    &     0.29    &      14.10    &      0.27    &     12.40   &       0.60    &    12.00   &      0.57   &    10.80   &    0.52   &   15.20   &    0.73   &   79.9    &    7.3   &      ...    &     ...   \\
5    &  249.00   &     5.51   &   376.00    &    14.50    &     380.00    &     10.20    &    383.00   &      30.30    &   289.00   &     18.80   &   247.00   &   13.60   &  266.00   &   17.3   &  808.0   &   74.90   &    875   &    98   \\
6    &    2.19   &     0.08   &     2.90    &     0.10    &       2.48    &      0.10    &      1.61   &       0.08    &     1.60   &      0.08   &     1.78   &    0.10   &    7.72   &    0.38   &  117.0   &   10.8   &     ...    &     ...   \\
7    &   57.20   &     1.26   &    96.20    &     2.04    &      98.50    &      1.91    &    118.00   &       5.99    &   117.00   &      5.84   &   102.00   &    4.87   &  107.00   &    5.35   &  120.0   &   11.1   &      ...    &     ...   \\
8    &  113.00   &     2.49   &   172.00    &     4.11    &     158.00    &      3.64    &     92.10   &       4.58    &    61.50   &      3.04   &    44.70   &    2.14   &   51.30   &    2.47   &  345.0   &   32.0   &      ...    &     ...   \\
9    &    1.86   &     0.06   &     2.93    &     0.07    &       3.19    &      0.11    &      2.36   &       0.12    &     2.66   &      0.13   &     3.43   &    0.18   &    5.53   &    0.27   &   89.8   &    8.3   &      ...    &     ...  \\
10    &    0.33   &     0.04   &     0.64    &     0.07    &       1.45    &      0.09    &      3.94   &       0.20    &     5.96   &      0.29   &     8.20   &    0.40   &   10.10   &    0.48   &   17.2   &    1.6   &      ...    &     ...   \\
 \hline \noalign {\smallskip}
\end{tabular}\label{table:c2d}
\noindent \scriptsize{Comments:  Only the first 10 lines are shown. The full table is available online.} 
\end{center}
\end{table*}

\section{Sample selection and ALMA Observations}

\subsection{Sample Selection}\label{sample}

At a distance of 140 $\pm$ 10 pc (Ortiz-Leon et al. 2017, Canovas et al. in prep.),  Ophiuchus is the closest  star-forming region with at least $\sim$300  discs. 
The ODISEA sample is the full catalog of  297 Young Stellar Objects (YSOs) in Ophiuchus from the \emph{``Cores to Discs'' Spitzer} Legacy Program (Evans et al. 2009b).
YSOs are usually divided into different classes based on their spectral slopes,

\begin{equation}
 \alpha_{IR} = 
  {\frac {log(\lambda_1 F_{\lambda_{1}})  -  log(\lambda_0 F_{\lambda_{0}})}     {log(\lambda _{1})-  log(\lambda _{0})   }},
 \end{equation} 
 
where F$_\lambda$ is the flux density at $\lambda_1$ and $\lambda_0$,  corresponding to  $\sim$2 and $\sim$20 $\mu$m, respectively (Greene et al. 1994; Chen et al. 1995).  
Class~I sources have $\alpha_{IR}$ $>$ 0.3 and are typically associated with  very young objects deeply embedded in their envelopes. 
Flat spectrum sources  have  0.3 $>$ $\alpha_{IR}$   $>$  -0.3 and are less embedded systems, but presumably still retain some detectable emission from the envelope.
Class~II objects have    -0.3 $>$ $\alpha_{IR}$   $>$  -1.6 and are sources where the infrared excess arises almost exclusively from an optically thick circumstellar disc. 
Finally, Class~III sources have $\alpha_{IR}$ $<$ -1.6 and are more evolved systems with little or no disc emission up to 20 $\mu$m. 

Using these classes as a guide, the full ODISEA sample is divided into two sub-samples, as shown in Figure~\ref{fig:sample}.  Sample A contains all Class I and Flat Spectrum sources, specifically objects  with [K]-[24] $>$  6.75 mag, and 
Class~II sources with K $>$ 10 mag. Sample B contains all Class III discs (they all have weak but significant, $>$5-$\sigma$, IR excesses) and Class II sources fainter than 10 mag in K-band. 
The boundary in K-band was chosen so that the two subsamples have $\lesssim$150 targets, which allows to fit each subsample in a single Science Goal in the ALMA Observing Tool and maximizes the efficiency of the observations.    
Both samples, A and B,  were observed in ALMA Cycle-4, but the observations for Sample B were not successful due to an scripting problem at the observatory that prevented the proper cycling between targets and 
phase calibrators. As a result, only a few  targets from Sample B were observed and the observations did not pass the Quality Assessment performed by the observatory. 
Therefore, in this paper we focus on the 147 sources from Sample A, which are listed in Table~\ref{table:sample}. 
Table~\ref{table:c2d} shows the 2MASS and \emph{Spitzer} photometry from these 147 objects,  taken from the NASA/IPAC Infrared Science Archive.\footnote{http://irsa.ipac.caltech.edu/Missions/spitzer.html}
The ODISEA project was approved again for ALMA Cycle-5, and 
at the time of this writing, the observations for Sample B have started, but have not yet been completed. 
We emphasize that Samples A and B are not equivalent. Objects in Sample B are expected to be, on average,  significantly fainter at millimeter wavelengths since 1) Class II objects that are fainter in K-band tend to have lower (sub)stellar masses and  disc masses correlate with the mass of the central object (Andrews et al. 2013), and 2) Class III disc tend to have very low dust masses ($<$ 0.3 M$_{\oplus}$; Hardy et al. 2015).

\subsection{The completeness of the \emph{Spitzer} disc census}
The  spatial distribution of both Samples and A and B are shown in Figure~\ref{fig:spitzermap} overlaid in the \emph{Spitzer} map of the ``Cores to Discs" Legacy  Project (Evans et al. 2009b). 
The map covers a region with an area of  6.6 deg$^2$ in the sky. Within this region, the \emph{Spitzer} catalog has the following 90$\%$ completness limits: 0.018, 0.020, 0.066, 0.100, and 0.700 mJy at 
3.6, 4.5, 5.8, 8.0, and 24 $\mu$m. 
Since less than a Moon mass of warm dust (T $\sim$100-300 K) is needed to produce an optically thick excess emission above the stellar photosphere in the mid-infrared ($\sim$8 to 24 $\mu$m),   this \emph{Spitzer} catalog represents an essentially complete IR census of the optically thick disc population in the stellar mass regime and extends well into the substellar members (Allers et al. 2006).
The extinction at 24 and 8 $\mu$m is only 2.5 and 5$\%$ of the visual extinction, respectively.  Therefore, extinction does not affect significantly the completeness of the survey. 
In fact,  highly embedded targets tend to be very young Class I objects, where the accretion luminosity increases the mass sensitivity of the IR observations (Evans et al. 2009b).
Dunham et al. (2008) estimates the sensitivity of the ``Cores to Discs" survey to be 4$\times$10$^{-4}$ L$_{\odot}$  for embedded protostars at 140 pc, 
also well below the stellar/substellar boundary.  

For Class III sources with small IR excesses above the stellar photosphere, the sensitivity of the \emph{Spitzer} survey is mostly given by the photospheric fluxes at 24 $\mu$m of stars of different masses and ages. 
Wahhaj et al.  (2010) find  that the  ``Cores to Discs" survey can reach the stellar photospheres of 0.3-0.5 M$_{\odot}$ pre-main-sequence stars at 140 pc for ages in the 
1 to 3 Myr range.  Such discs typically have low luminosities (L$_{disc}$/L$_{star}$ $\lesssim$ 10$^{-3}$) and are optically thin at mid-IR wavelengths. 
Hardy et al. (2015) observed 24 \emph{Spitzer}-selected Class III discs  with ALMA using a sensitivity similar to that of ODISEA. They detected only 4 targets in the continuum and none of them in CO, suggesting 
that these Class III objects are in a very advanced stage of disc dispersal or already in the debris disc phase. 
Therefore, while some Class IIII discs around very-low mass stars ($\lesssim$ 0.3-0.5 M$_{\odot}$)  might be missing in the \emph{Spitzer} catalogs, those objects are unlikely to be detected by a snapshot survey like ODISEA. 
In principle, \emph{Spizer} observations could miss discs with no mid-IR excesses ($\lambda$ $\sim$8-24 $\mu$m), but significant far-IR ($\lambda$ $\sim$70-250 $\mu$m) emission. However,  far-IR surveys with \emph{Herschel} show that those systems are very rare and are consistent with cold debris discs or background galaxies (Cieza et. 2013;  G{\'a}sp{\'a}r \& Rieke, 2014; Rebollido et al. 2015).
In summary, the  \emph{Spitzer} disc census in the area mapped by the  ``Cores to Discs"  project  can be considered to be complete in the stellar mass regime for Class I and Class II sources, but might become incomplete for Class III sources, specially around very low-mass stars ($\lesssim$ 0.3-0.5 M$_{\odot}$). 

\subsection{Observations and data reduction}

All of our 147  targets in Sample A were observed under the Cycle-4 ALMA program 2016.1.00545.S  
on a single scheduling block, which was executed 3 times between July 13th and 14th 2017\footnote{
We note that previous observations of Sample A suffered from the same scripting problem as Sample B and are not included in this paper. The failed observations for Sample A correspond to  execution blocks uid://A002/Xc1c1f1/X81bc (July 7th, 2017) 
and uid://A002/Xc1e2be/X717 (July 9th, 2017). The failed observations for 
Sample B correspond to execution blocks uid://A002/Xbfb22d/X1b7f (April 27th, 2017) and uid://A002/Xbfb22d/X84ed (April 28th, 2017).}
The nominal array configuration was C40-5, and 42-45 of the 12-m ALMA antennas were used
with baselines ranging from 17 to 2647 m.
The precipitable water vapor (PWV) ranged from 1.1 to 1.9 mm  during the observations.
The  objects J1517-2422 and J1733-1304 were observed as flux calibrators, while the quasars J1517-2422 and J1625-2527 were 
used as bandpass and phase calibration respectively. 
The ALMA correlator setup was the following: three spectral windows 
were centered at at 230.538000, 220.398684, and  219.560358 GHz to cover the  $^{12}$CO J = 2-1,   $^{13}$CO J = 2-1, and  C$^{18}$O J = 2-1 transitions, respectively.  
All three windows had a spectral resolution of 0.08 km s$^{-1}$. The first spectral window had a bandwidth of 117 MHz, while the other two had 58.6 MHz bandwidths.
Two additional spectral windows, centered at 233.00 and  218.00 GHz,  had 1.875 GHz bandwidths and were selected for continuum observations, for a total continuum bandwidth of 3.98 GHz.

All data were calibrated  using the Common Astronomy Software Applications package (CASA v4.4;  McMullin et al. 2007) by the ALMA observatory. 
The standard calibration  included offline Water Vapor Radiometer (WVR) calibration, system temperature correction, bandpass, phase and amplitude calibrations. 
The observations from all three nights were concatenated and processed together to increase the signal to
noise and \textit{uv}-coverage. 
The flux calibration in the three epochs agreed  within $<$10$\%$ and thus no rescaling of the flux was applied. 
We used the CLEAN algorithm to image the data adopting Briggs weights and  robust parameter equal to zero  as a balance between resolution and sensitivity. 
For the continuum, we obtained a typical  rms of $\sim$0.15 mJy beam$^{-1}$ and a synthesized beam of 0.28$''$ $\times$ 0.19$''$. 
The molecular line data ($^{12}$CO, $^{13}$CO, and C$^{18}$O)  will be discussed in another paper of the series (Williams et al. in preparation). 

\section{Results}

\subsection{Continuum images, disc photometry, and sizes}\label{continuum}

We used the Viewer task within CASA to individually inspect all the images. In the majority of the cases, we found a single 1.3 mm detection within $<$1$''$ of the nominal location of the \emph{Spitzer} source that can be unambiguously identified as the target\footnote{the only exception object 62, where a 180 mJy detection is displaced 5$''$ arcseconds with respect to the \emph{Spitzer} coordinates. 
The detection correspond to the the object EM * SR 24S , which is part of a  triple system that  also contains the SR 24 Nb and SR 24Nc  components. We note that the \emph{Spitzer} 
coordinates correspond to the northern pair, which is only 0.34 mJy at 1.3 mm (Fernandez-Lopez et al. 2018).}.  
We first used the \emph{imstat} task to search for the peak flux value within an aperture of 1$''$ in radius and calculated an rms from an annulus with an inner and outer radius of 1.0$''$ and 1.2$''$. 
For \emph{single} objects with peak signal to noise  ratios (S/N) $>$ 5, we used the \emph{uvmodelfit} task in CASA  to derive basic parameters for each source.
The full spectral coverage was utilized. 
We fitted both a point source and Gaussian with all free parameters: the  integrated flux density,  the Full Width at Half Maximum (FWHM) along the major  and minor axis, the Position Angle (PA), and small offsets ($<$ 1.0$"$) in right ascension and declination from the phase center.
We find that for objects with peak S/N $\lesssim$ 30, the point source and the Gaussian fits give similar fluxes, but the Gaussian fits produce FWHM and PA values with very large uncertainties
($\gtrsim$ 50$\%$ and $\gtrsim$90 deg, respectively).  For objects with high S/N,  the Gaussian fit can measure FWHM values (deconvolved from the beam) down to a factor of $\sim$2 of the beam size, but the ability to measure the FWHM and the PA of the disc depends on both the size and the S/N of the source. 
For sources with sizes (i.e.,  FWHM values of major axes) comparable to the beam, the Gaussian and point source fit give similar fluxes, although the Gaussian fit tends to give slightly larger fluxes (by $\sim$10$\%$). 
For more extended sources, the point source fit significantly underestimates the flux.

Ideally, one would use the ratio of the FWHM value to its uncertainty to establish whether a source is spatially resolved. However, the \emph{uvmodelfit}  
documentation indicates that the size uncertainties are likely to be underestimated by this task. 
Therefore, we use  \emph{ad hoc} criteria to decide whether the source is spatially resolved (i.e., sufficiently different from a point source to justify reporting disc sizes and orientations). 
In particular, we provide size information only  if the source has a peak S/N $>$ 30 and the Gaussian flux is greater than the point source flux plus 3 times the rms of the sky between 1 and 1.2$''$ of the target. 
For resolved sources, we report the 1-$\sigma$ uncertainties obtained for each parameter (except for the position), 
but remind the reader that these uncertainties might be underestimated.  
Otherwise, we consider the source to be unresolved and  only report the flux and its uncertainty.  
For the few objects with significant substructures  (e.g., transition discs objects with cavities, spiral arms and/or wide gaps,  targets \# 12, 22A, 51, 41, 62, 127, 141, and 143, see Section~\ref{structures}), the Gaussian model does not provide an accurate fit. In these cases, we fit an elliptical disc model to measure their sizes  (also within \emph{uvmodelfit}), but we use the 2-D fitting tool within the CASA Viewer to measure the flux using a sufficiently large aperture, typically  $\sim$3$''$ in radius.  

For multiple sources (doubles and triples) we also measure fluxes and discs sizes (also expressed in terms of the FWHM of the major axes) in the image plane using the 2-D fitting tool within CASA. This tool provides size and PA information for spatially resolved objects that have enough S/N.  Otherwise, the task indicates that the source is consistent with a point source.  For these multiple sources,  we use the convergence of the fitting tool as the detection criteria of the sources.

For non-detected targets, we still estimate and report the flux and rms at the expected location of the source using  \emph{uvmodelfit} to fit a point source. From all the detections, we find average offsets in RA and Dec of -0.08$"$ and -0.56$''$, respectively, which we attribute to the proper motions of the targets based on the results of the Gaia Data Release 2 (Gaia Collaboration et al. 2018).
Canovas et al. (in preparation) found  Gaia proper motions of $\mu_\alpha$ =   -6.9 $\pm$ 1.6 and $\mu_\delta$ = -25.6 $\pm$ 1.7 $''$/year for a sample of $\sim$200 Ophiuchus discs (most, but not necessarily all, 
are included in our  ALMA sample).  Since the \emph{Spitzer} catalogs from the  ``Cores to Discs"  project (Evans et al. 2009b)  used to select our targets were tied to 2MASS coordinates (Cutri et al. 2003), there is a $\sim$18 year difference between the 
\emph{Spitzer} coordinates  and the ALMA observations.  This should translate to a  shift in coordinates of $\sim$0.1$''$ and $\sim$0.5$''$ in RA and Dec, respectively, consistent with the observed offsets. 
Because the mean displacement is larger than the beam,  we apply the observed offsets to the nominal position of the non-detected targets before fitting a point source at the new location.  

Given the different methods used to estimate the photometry and the disc sizes, we emphasize that the fluxes should be taken with caution, especially for objects with clear substructures.  We also note that all the fluxes are subject to a 10$\%$ calibration uncertainty. 
All the information on disc sizes and fluxes, obtained as described above is listed in Table~\ref{table:results}. 

\begin{figure*}
\includegraphics[width=18cm, trim = 0mm 0mm 0mm 0mm, clip]{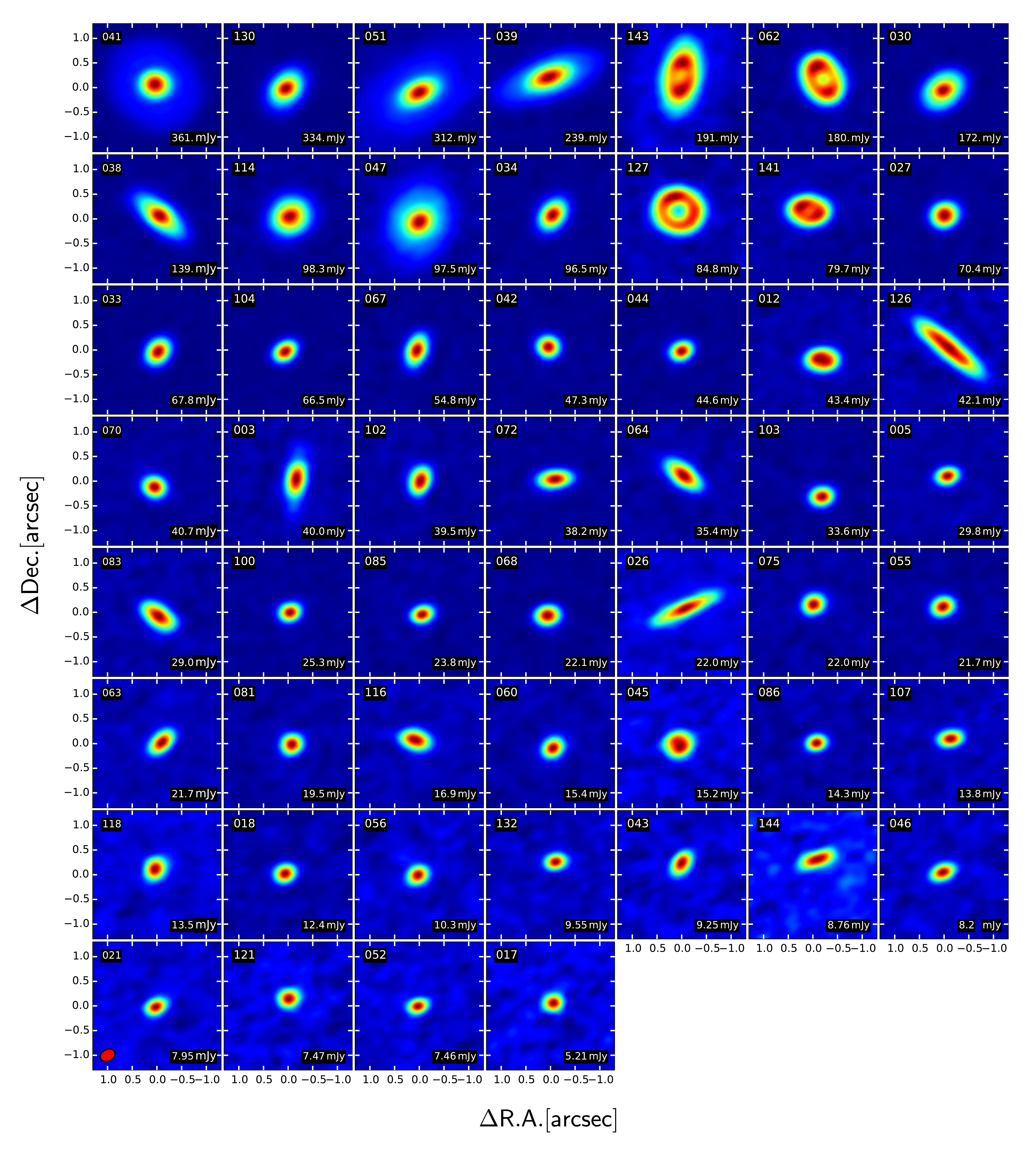}
\caption{ \small The 53  individual detections resolved by our 1.3 mm observations,  ordered by decreasing integrated flux (provided at the bottom right of each panel).
}
\label{fig:resolved}
\end{figure*}

\begin{figure*}
\includegraphics[width=18cm, trim = 0mm 0mm 0mm 0mm, clip]{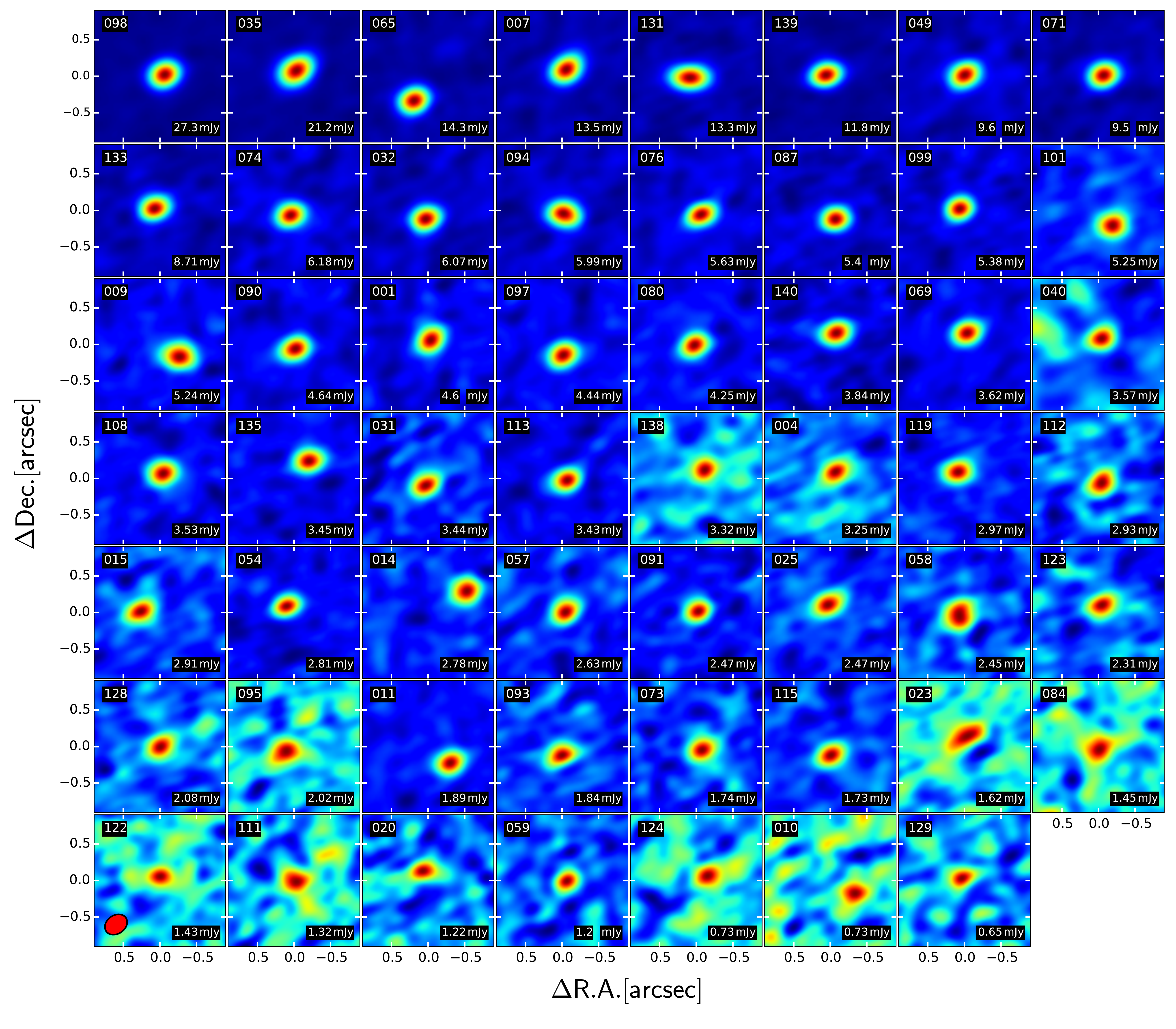}
\caption{ \small The 55 individual detections that remain unresolved in our 1.3 mm data, also ordered by decreasing integrated flux (provided at the bottom right of each panel). 
}
\label{fig:point}
\end{figure*}

\begin{figure*}
\includegraphics[width=18cm, trim = 15mm 10mm 0mm 15mm, clip]{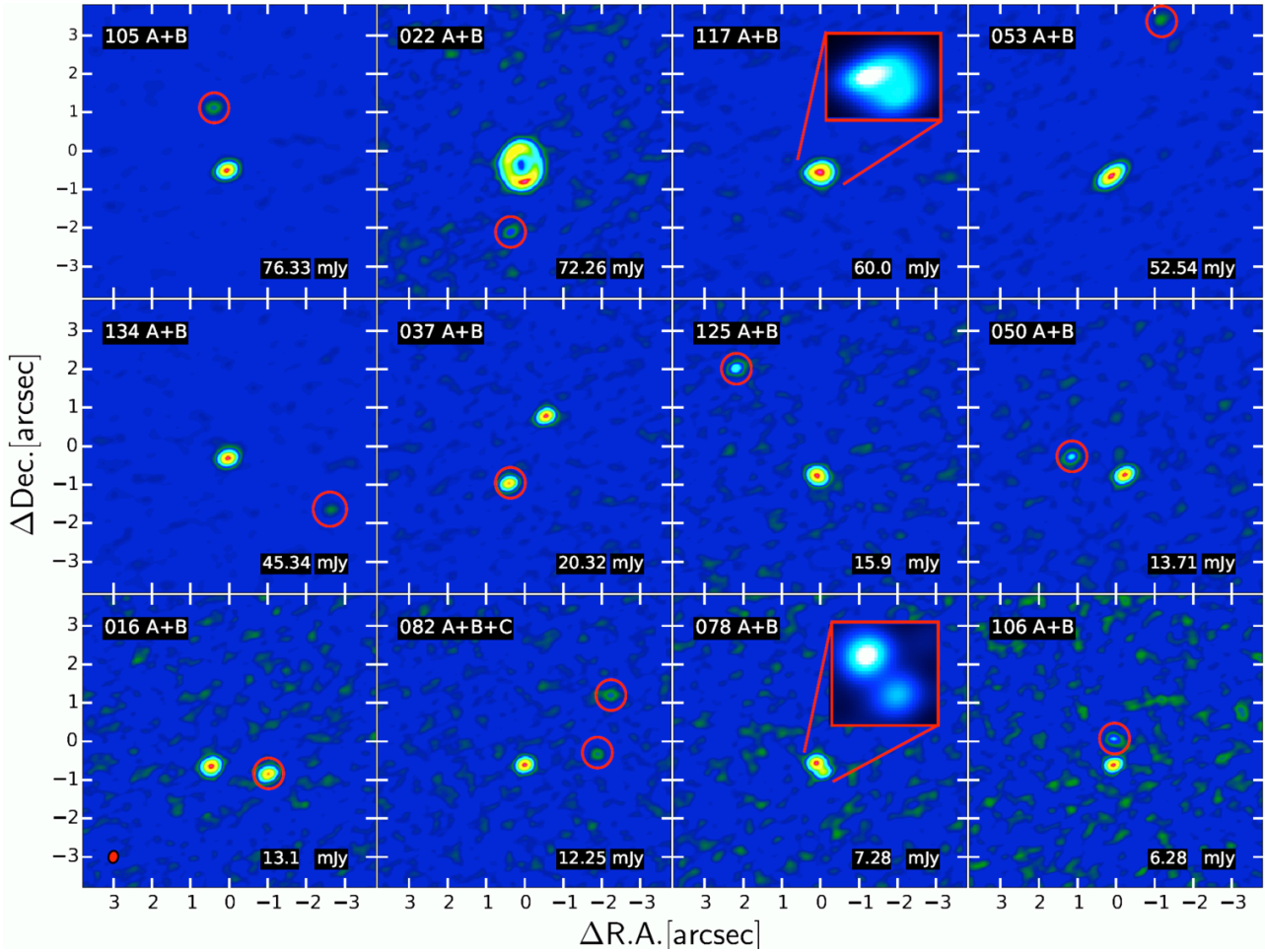}
\caption{ \small The 12  multiple systems in which discs are seen around each one of the individual components.
The fluxes listed correspond to the primaries. 
Object 082 is a triple system, the rest are binaries. Object  117 is a known infrared binary which is only barely resolved by our ALMA observations. 
Companions are indicated with red circles or highlighted in insets of images with higher spatial resolution (i.e., uniform weights were applied in cleaning). 
}
\label{fig:binaries}
\end{figure*}

We detect a total of 133 discs in 120 systems. These 133 discs correspond to 108 single discs plus the individual components of 11 binary discs and 1 triple disc system, leaving 27 targets undetected. 
Figure~\ref{fig:resolved} shows the 53 single discs with size information ordered by decreasing integrated flux. Many of the brightest sources are clearly extended, while  most of the fainter sources tend to be only partially resolved. 
Figure~\ref{fig:point} shows the 55  single discs that are detected but are consistent with a point source at the resolution of our observations ($\sim$0.2$''$).  The vast majority (49/55) of these sources are fainter than $\sim$10 mJy.
Multiple targets (11 binaries and 1 triple system) are highlighted in Figure~\ref{fig:binaries}. 
In Table~\ref{table:results} we add an ``A" to the source ID to denote the component that is closer to the nominal coordinates from \emph{Spitzer}  (i.e., the center of the ALMA pointing) and a ``B" to denote the other source. In the case of the triple system, the source farthest away from the ALMA pointing is defined as the ``C" component. 

\begin{table*}
\begin{center}
\caption[]{\em{Fluxes and sizes at 1.3 mm from Cycle-4}}
\begin{tabular}{@{}lccccccccccc@{}}
\hline
ODISEA C4 ID         &    RA           &   Dec        &    F$_{1.3}$  &  eF$_{1.3}$ & Major  & eMajor&  Minor &  eMinor  &   PA  & ePA  \\
                                    &  (J2000.0) & (J2000.0) &  (mJy)                  &     (mJy)   & (mas)  & (mas) &       (mas)  & (mas) &          (deg)  & (deg)   \\
\hline
ODISEA\_C4\_001  &  16:21:31.924 &  -23:01:40.79   & 4.60  &  0.16  &   ... &     ...  &    ...  &    ...  &    ...  &    ...  \\
ODISEA\_C4\_002  &   ...  &  ...   & 0.17  & 0.20   &    ...  &    ...  &     ...  &    ...  &    ... &    ...   \\
ODISEA\_C4\_003  &16:21:45.123   & -23:42:32.19  & 40.09  & 0.23  &  541.0   &  3.6  &   97.2 &    5.0   & 174.2  &    1.0 \\
ODISEA\_C4\_004  & 16:21:48.474  & -23:40:27.76   & 3.25  & 0.31   &  ...  &    ...  &    ... &    ...  &   ...  &   ... \\
ODISEA\_C4\_005  & 16:22:18.522  & -23:21:48.62  & 29.84  & 0.24  & 184.4   &  4.9 &  133.2   &  6.0  & 108.7  &   0.3 \\ 
ODISEA\_C4\_006  &  ...  &  ... & -0.19  & 0.13   & ...  &   ...   &  ...   &   ...   &  ...  &    ...  \\
ODISEA\_C4\_007  & 16:22:24.953  & -23:29:55.41  & 13.57  & 0.18   & ...  &   ...   &  ...   &  ...   & ...   &  ... \\
ODISEA\_C4\_008  &  ... &  ... & -0.03  &  0.31  &  ...   &  ...   &  ...    & ...   &  ...   &  ...  \\
ODISEA\_C4\_009  & 16:23:05.418  & -23:02:57.49  &  5.24  & 0.19   & ...    & ...   & ...      & ...   &  ...   &  ... \\
ODISEA\_C4\_010  &  ... &  ... &   0.73 &  0.13  &   ...   &  ... &     ...   &  ...  &    ...   &  ... \\
\hline
\end{tabular}\label{table:results}
\noindent 

\scriptsize{Comments:  Only the first 10 lines are shown. The full table is available online. The ODISEA\_C4 designation indicates that the data were obtained in ALMA Cycle-4. 
The next number denotes  the corresponding target  in Table 1. 
Sources with no RA and Dec information have not been detected by our ALMA observations.
Sources with position information but no size information have been detected but remain unresolved.  
The major and minor axis information correspond to FWHM values.}
\end{center}
\end{table*}

\subsection{Stacking of non-detections}\label{stacking}

\begin{figure*}
\includegraphics[width=8cm, trim = 0mm 0mm 0mm 0mm, clip]{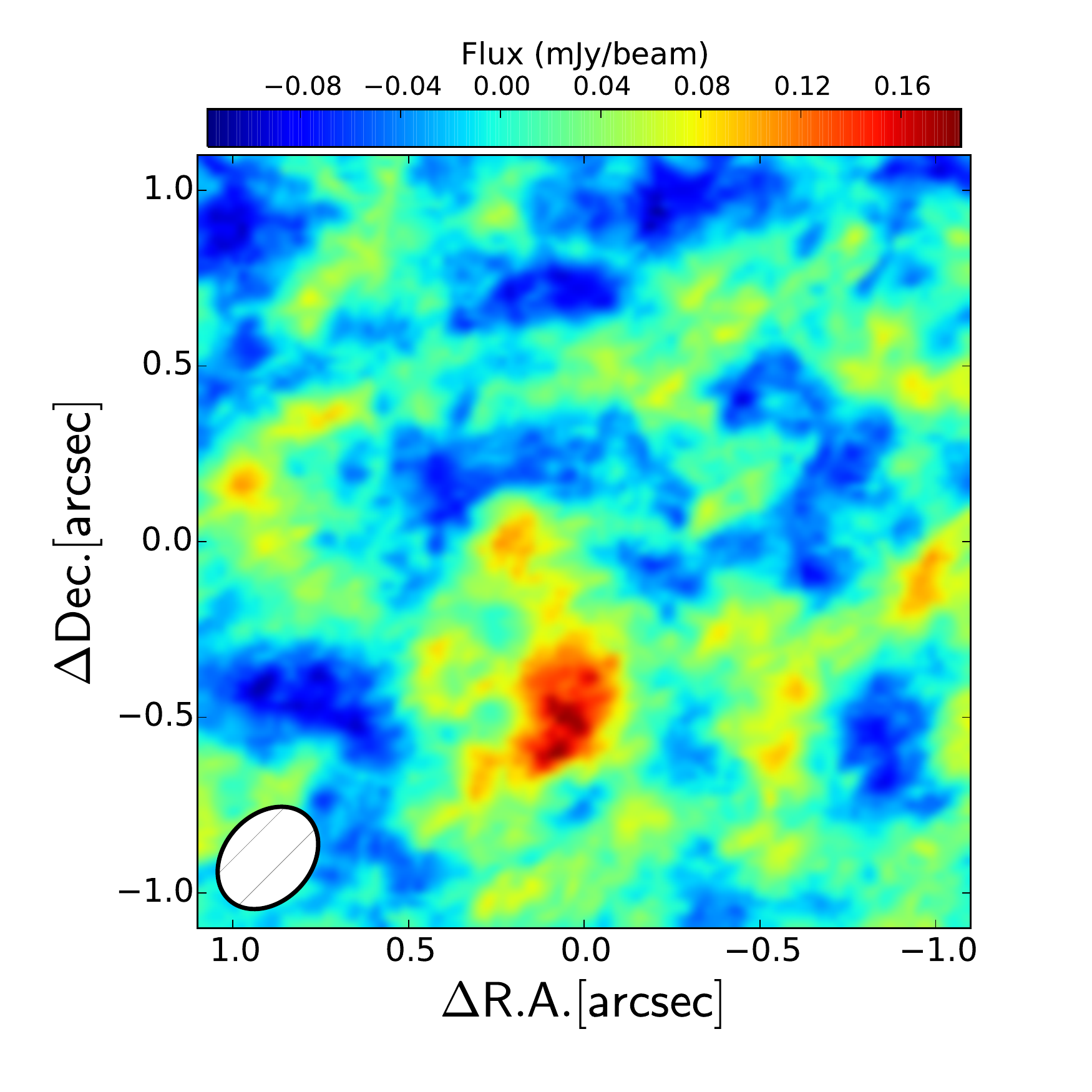}
\caption{ \small Stacking of the 27 non-detections showing a 0.16 mJy (4-$\sigma$) detection with an offset  of -0.5$''$ in RA,  consistent with the mean RA offset seen in the detected discs.}
\label{fig:stacking}
\end{figure*}

To estimate the typical flux of the discs that were not detected, we stack the images of the 27 non-detections, calculating the mean of each pixel from the individual images 
centered at the nominal location of each target. The results are seen in Figure~\ref{fig:stacking}.
The stacked image has an rms of 0.04 mJy and shows a 4-$\sigma$ (0.16 mJy) detection shifted -0.5$''$ in RA. This offset is
very consistent with the mean offset in RA found for the detected discs (-0.56$''$) and suggest that the detection is real and probably diluted by the dispersion in the  individual offsets
of the targets included.     
This implies that there are several discs in our sample with fluxes close 
to the 1-$\sigma$ noise of the individual observations. 

\subsection{General trends}

\begin{figure*}
\includegraphics[width=10cm, trim = 0mm 80mm 0mm 0mm, clip]{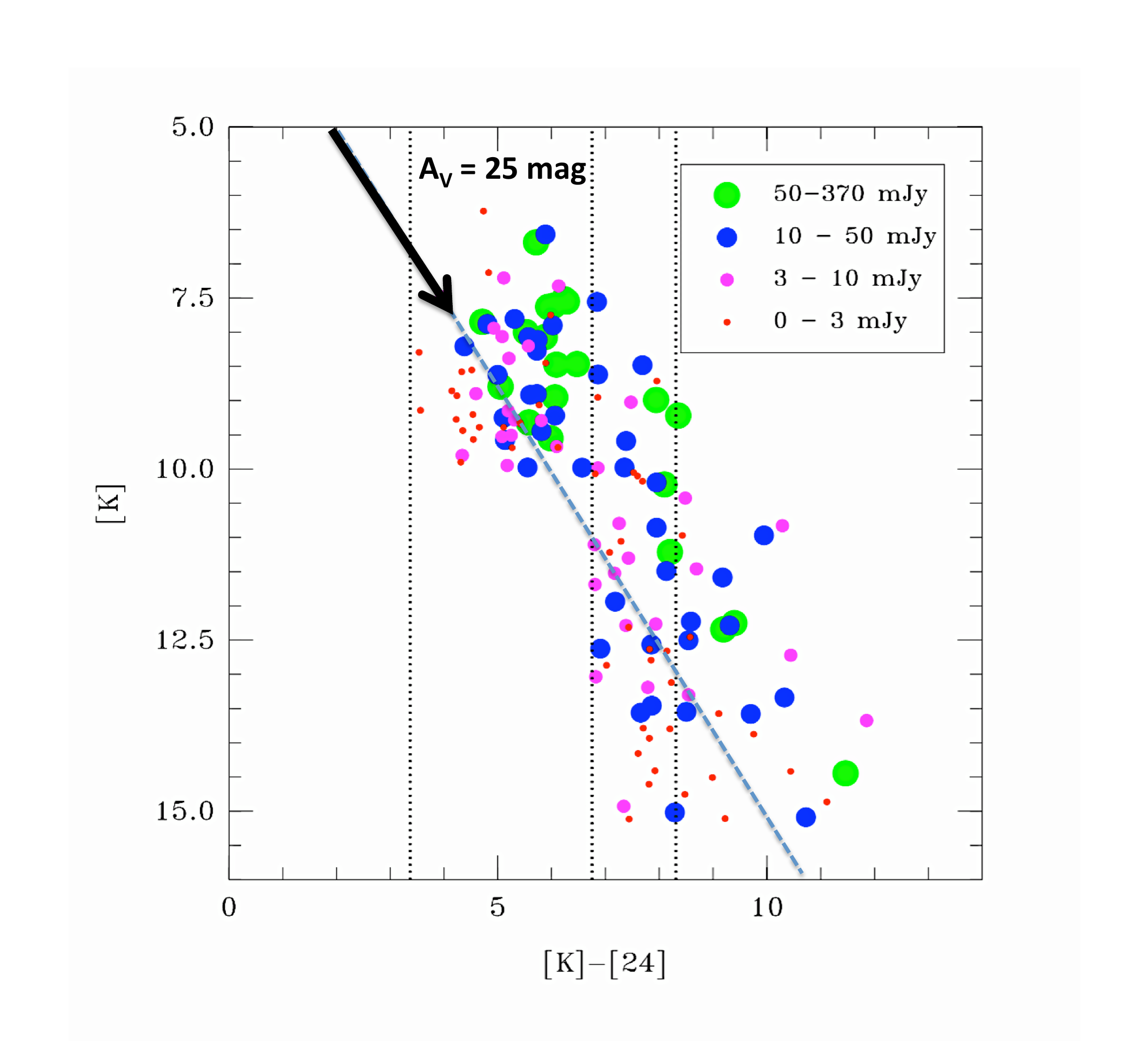}
\caption{ \small The K-band magnitude as a function of [K]--[24] color for our Cycle-4 sample, with the sources color-coded by
flux. ALMA non-detections are included on the lower bin (0 to 3 mJy).  
The black arrow represent a  A$_V$ = 25 mag  extinction vector. 
As in Figure 1, the vertical dotted lines are the approximate boundaries between Class III, Class II, Flat Spectrum, and Class I objects (left to right)
Brighter sources tend to occupy the  upper envelope of the [K] vs [K]--[24] plane (above the dashed diagonal line) and correspond to the brightest infrared sources and/or the most embedded objects. 
}
\label{fig:colormag-detec}
\end{figure*}

\begin{figure*}
\includegraphics[width=8.5cm, trim =  0mm  0mm 0mm 0mm, clip]{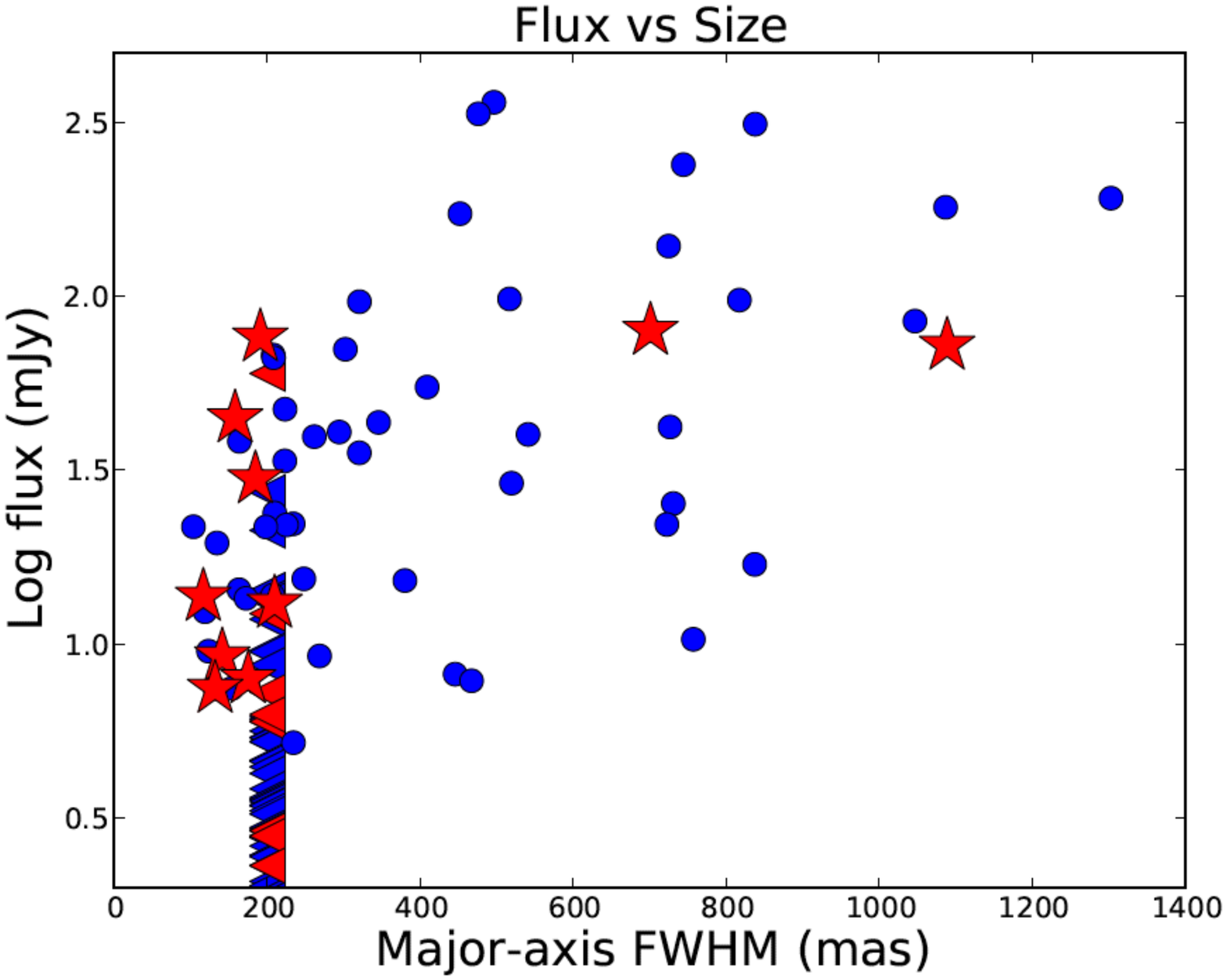}
\includegraphics[width=8.5cm, trim =  0mm 0mm  0mm 0mm, clip]{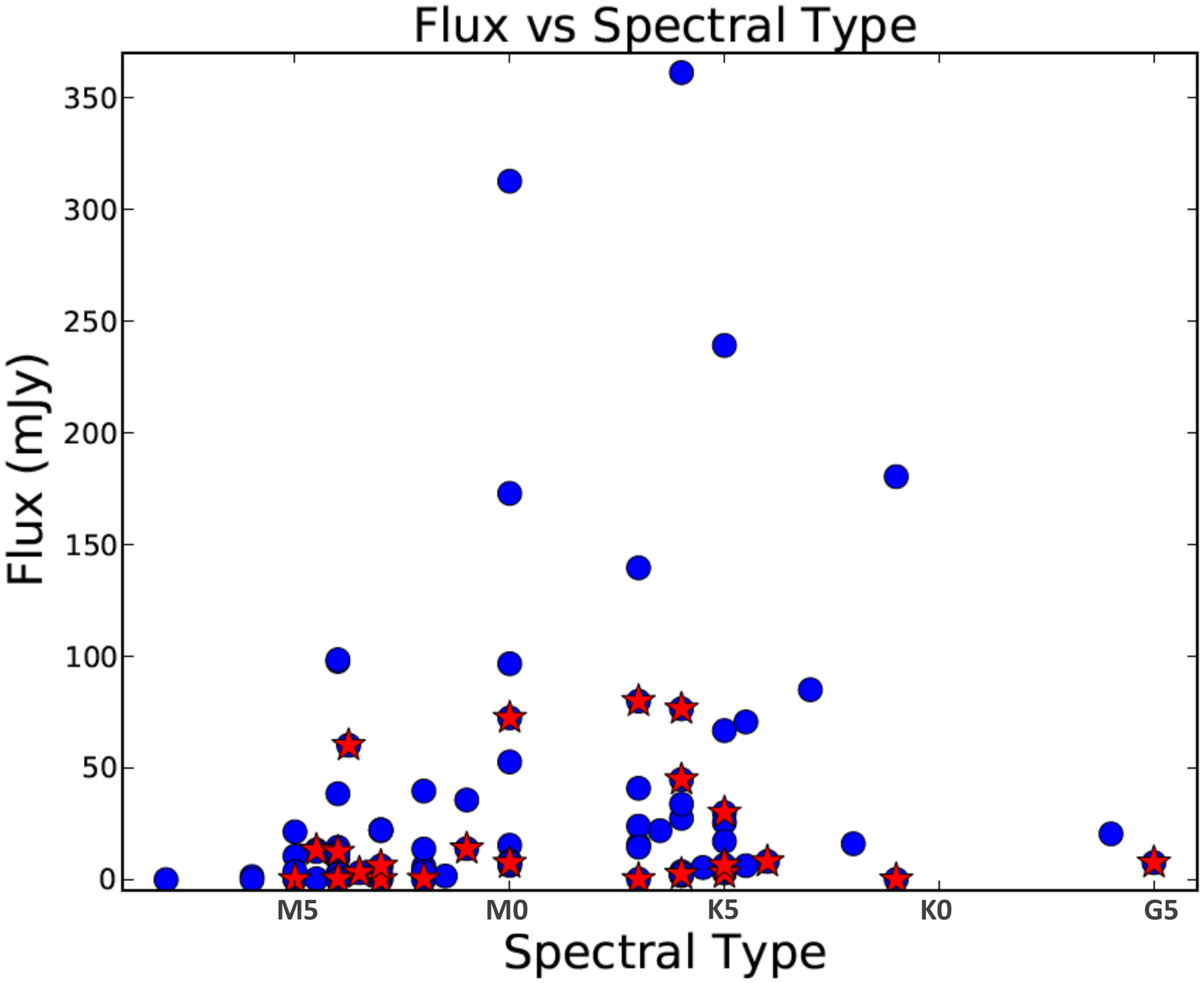}
\caption{ \small 
{\bf{Left Panel:}} the 1.3 mm flux vs the FWHM of the detected discs. 
Members of multiple systems are shown in red. Unresolved sources (triangles) are shown with upper limits in disc sizes of 200 mas and 
are mostly clustered toward low fluxes ($\lesssim$ 10 mJy). Discs in multiple systems tend to be small, with a few exceptions (objects number 022A and  141). 
While there is a large dispersion of sizes  for a given flux, there is a general trend in the sense that brighter discs tend to be larger.  
{\bf{Right Panel:}} the 1.3 mm flux of the disc vs the spectral type of the central object. 
Objects with close companions (projected separation  $<$ 2.0$''$) are indicated by red stars.  
Faint discs are seen around  almost all types of stars,  but the brightest discs (F$_{1.3mm}$ $>$ 100 mJy) are mostly restricted to objects with types between M0 and K5 without known close companions.
}
\label{fig:fluxsize}
\end{figure*}

In Figure~\ref{fig:colormag-detec} we show our detections  in the infrared color-magnitude diagram used to divide the ODISEA sample into two (see Figure~\ref{fig:sample}).
The detections are color-coded based on their observed 1.3 mm flux. 
The brightest millimeter sources occupy the upper envelope of the [K] vs [K]-[24] plane, which represent the brightest near-IR sources and/or the most embedded objects, which in turn 
tend to be embedded Class I sources and Class II discs around (relatively) more massive stars. 
Similarly,  the fainter millimeter sources  ($<$ 3 mJy) tend to have fainter K-band fluxes and bluer [K]-[24] colors. 
Therefore, it is reasonable to suspect that most of the ODISEA targets that still remain to be observed (Sample B in Figure~\ref{fig:sample}) will be, on average,   fainter at 1.3 mm. 
Also, the A$_V$ = 25  mag extinction vector shown at the top left of the figure demonstrates that  very high extinction can move sources across the IR SED classification (e.g., from Class III to Class II and from Class II to Flat Spectrum and Class I sources).  This introduces some ambiguity in the physical interpretation of the SEDs Classes,  which are purely based on the observed spectral slopes. 
For instance, while there is a strong correspondence between observationally defined  Class I sources and theoretical Stage I objects (protostars surrounded by infalling envelopes;  see Evans et al. 2009a), young stellar objects without associated envelopes  might also be classified as Class I sources if they are  found behind a dense molecular cloud. 
We have searched the literature for multiplicity information and found that 25 of our targets have previously known companions within 5$''$. Our ALMA imaging survey has identified  4 additional binaries, for 
a total of 29 multiple systems among our 147 ALMA targets. 
We have  also collected spectral types from the literature for 88 objects from our sample. The spectral types and the separations of the companions are listed in Table 1. 
In Figure~\ref{fig:fluxsize} (left panel) we show the flux as a function of major axis for our sample. There is a general trend in the sense that brighter discs tend to be larger as already reported by Pietu et al. (2014) and 
Tripathi et al.  (2017).
However, we note that there is a very large dispersion of sizes for a given flux. For instance, discs with $\sim$50 mJy fluxes have FWHM values that range from smaller than $\sim$0.1$''$ to $\sim$1$''$,  suggesting very different surface density profiles in the dust.
Unresolved sources are typically faint (F$_{1.3,}$ $\lesssim$ 10 mJy). 
Discs with close companions (projected separations $<$ 2.0$''$; shown with red symbols)  tend to be small (FWHM $<$ 0.2$''$ or 28 au), with the exception of objects  022A  and 141. Object 022A (see Figure~\ref{fig:binaries}) has a relatively distant companion (1.78$''$ or 250 au), while object 141 has recently been identified as a circumbinary disc using aperture masking imaging (Ruiz-Rodriguez et al. 2016). In this latter case,
the separation of the companion is only 20 mas (2.8 au). 
Stellar companions are known to decrease the incidence of circumstellar discs in young stars (Cieza et al. 2009; Kraus et al. 2012), specially in systems with separations of a few tens of au. 
Cox et al. (2017) recently demonstrared that  discs in binary systems are also smaller and fainter than those around single stars, which is consistent with our results shown in Figure~\ref{fig:fluxsize}.  
A detailed discussion of the effects of stellar companion on the properties of protoplanetary discs will be included in a future paper of this series (Zurlo et al. in preparation), which also presents a dedicated adaptive optics 
search for stellar binaries in Ophiuchus. 
In Figure~\ref{fig:fluxsize} (right panel), we also  show the disc flux as a function of spectral type, which serves as a rough proxy for stellar mass. 
Disc properties as a function of stellar properties will be quantified and  investigated in more detail in a  follow-up paper from this series incorporating new optical and infrared spectroscopic observations  (Ruiz-Rodriguez et al. in preparation). 
However,  in our limited sample, we find that the fainter discs (F$_{1.3mm}$ $<$ 20 mJy) are distributed across all spectral types, while 
the brighter discs (F$_{1.3mm}$ $>$ 100 mJy) are clustered around spectral types in the K5 to M0 range in objects without known stellar companions.  

In summary, from  the initial analysis of the ODISEA sample we can report the following general trends:   faint (F$_{1.3mm}$ $<$ 10 mJy) and small (FWHM $<$0.2$''$ $\sim$28 au) discs are the most common type of discs in the Ophiuchus Molecular Cloud. 
They are seen across spectral types and SED Classes, and specially around stars in binary systems.  Discs brighter than 50 mJy  and 100 mJy represent only $\sim$15$\%$ and $\sim$5$\%$ of our sample of 147 objects, respectively. 
Given the trends seen in Figure~\ref{fig:colormag-detec}, which suggest that bright discs are already over-represented in ``Sample A" with respect to ``Sample B", 
these very bright discs most likely account for  even smaller fractions of the full population of \emph{Spitzer}-detected discs.  
The brightest discs in our sample (F$_{1.3mm}$ $>$ 100 mJy) are seen around  (presumable) single stars with intermediate spectral types and seem to be missing at the edges of the spectral type distribution.  
There is a tendency for brighter discs to be larger, but the flux vs size relation show a significant dispersion (a factor of $\sim$10).  

\subsection{Disc substructures}\label{structures}

\begin{figure*}
\includegraphics[width=15cm, trim = 0mm 0mm 0mm 0mm, clip]{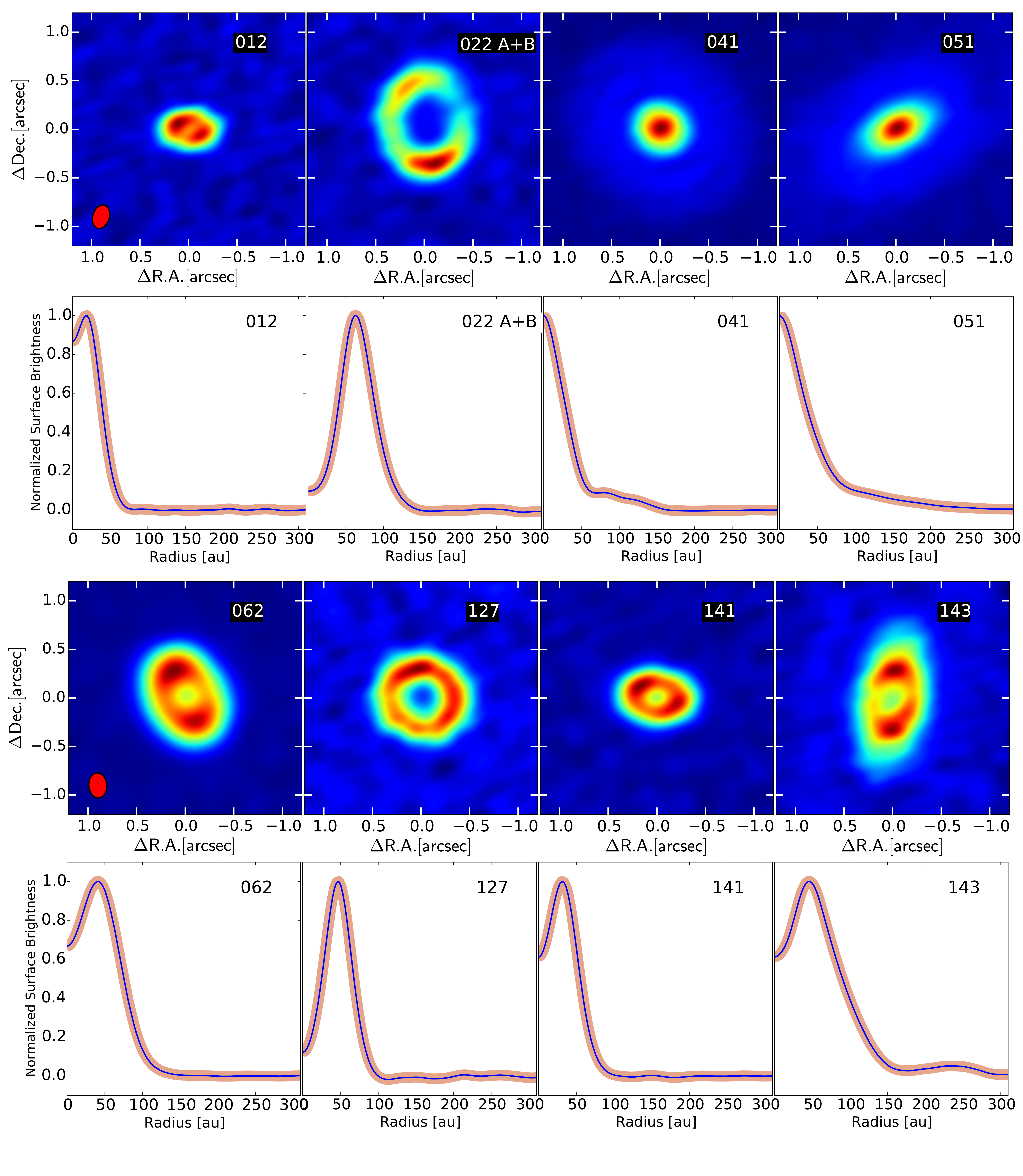}
\caption{ \small Images (left) and radial profiles (right) of the ODISEA objects with clear substructures: inner cavities and/or narrow gaps.
The red shade represent the 1-$\sigma$ error on the mean, 
given by the dispersion at a given radius divided by the square root of beams (i.e. the number of independent elements).
The  blue shade corresponds to the dispersion in azimuth at each radius. 
}
\label{fig:structures}
\end{figure*}

For the subset of  $\sim$50 sources with  fluxes $\gtrsim$ 15 mJy,  we perform phase-only self-calibration and produced images with \textit{uniform} weightings to increase 
the resolution from $\sim$0.2$''$ to $\sim$0.15$''$ and search for substructures in the discs. 
For this subsample, the average improvement in the peak signal-to-noise ratio (S/N) is 10$\%$, but the improvement
can be as high as $\sim$50$\%$ for the brightest sources. 
Below a flux level of 15 mJy, self-calibration does not significantly improve the S/N.  
A visual inspection of our images reveal a series of interesting substructures (see Figure~\ref{fig:structures}). 
Objects number 12, 22A,  62, 127, 141, and 143 show inner cavities of different sizes from barely resolved (object 12) to 1$''$ in diameter (object 22). 
Object 41 shows two concentric gaps at 0.4, and 0.7$''$. Object 143 shows an external ring in addition to the inner cavity. 
Objects 62, 127, 141 (best known as EM$^{\star}$ SR 24S,  DoAr 44, and RX J1633.9-2442 , respectively) had their cavities resolved by pre-ALMA observations (Andrews et al. 2011; Cieza et al. 2012b).
The substructures in objects 12 and 143 (also known as IRAS 16201-2410 and WSB 82) were recently imaged by Cox et al. (2017).  
Object 41 (best known as Elias 2-24) was recently identified as a disc with remarkable substructure almost simultaneously by three different groups (Cieza et al. 2017; Cox et al. 2017; Dipierro, G. et al. 2018).  
Object 51 (Elias 2-27) has spiral arms that were first identified by Perez et al. (2016).  
Interestingly, object 22A (also known as ROXRA 3) has the disc with the largest cavity in our entire sample but was not previously known to host such a cavity.

We note that all 8 targets discussed above showing structures are brighter than $\sim$43 mJy and therefore are among the brightest 25 objects ($\sim$17$\%$ of the sample). 
This implies that $\sim$28$\%$ (7/25) of the brightest objects show some type of substructures when observed at $\sim$0.2$''$ ($\sim$28 au) resolution. 
Whether the fainter (usually smaller) targets show scalled-down versions of these substructures remains to be established by deeper and higher-resolution observations. 

\begin{figure*}
\includegraphics[width=15cm, trim = 0mm 0mm 0mm 0mm, clip]{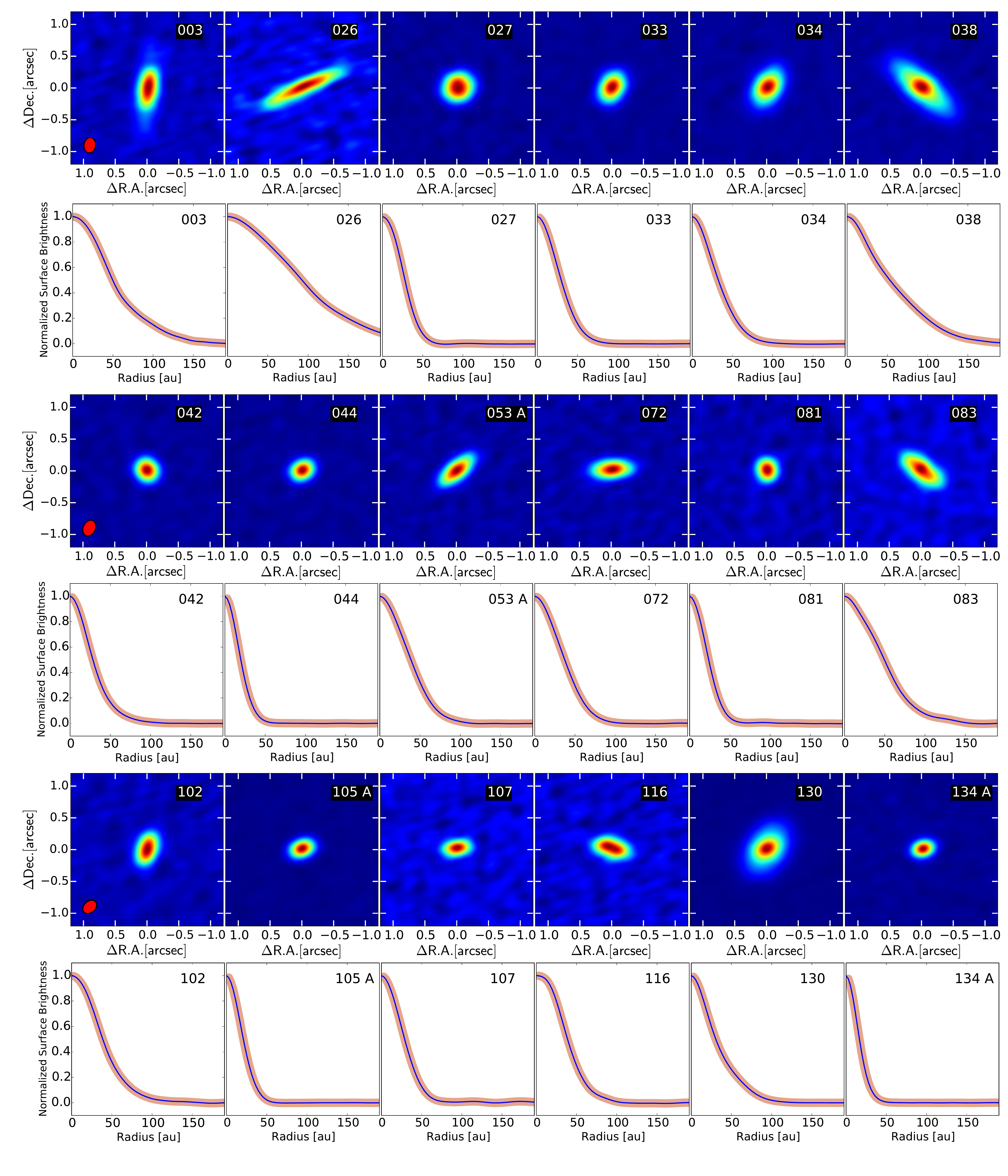}
\caption{ \small  Images (left) and radial profiles (right) of the ODISEA objects consistent with continuous discs. 
}
\label{fig:smooth1}
\end{figure*}

\begin{figure*}
\includegraphics[width=18cm, trim = 0mm 40mm 0mm 20mm, clip]{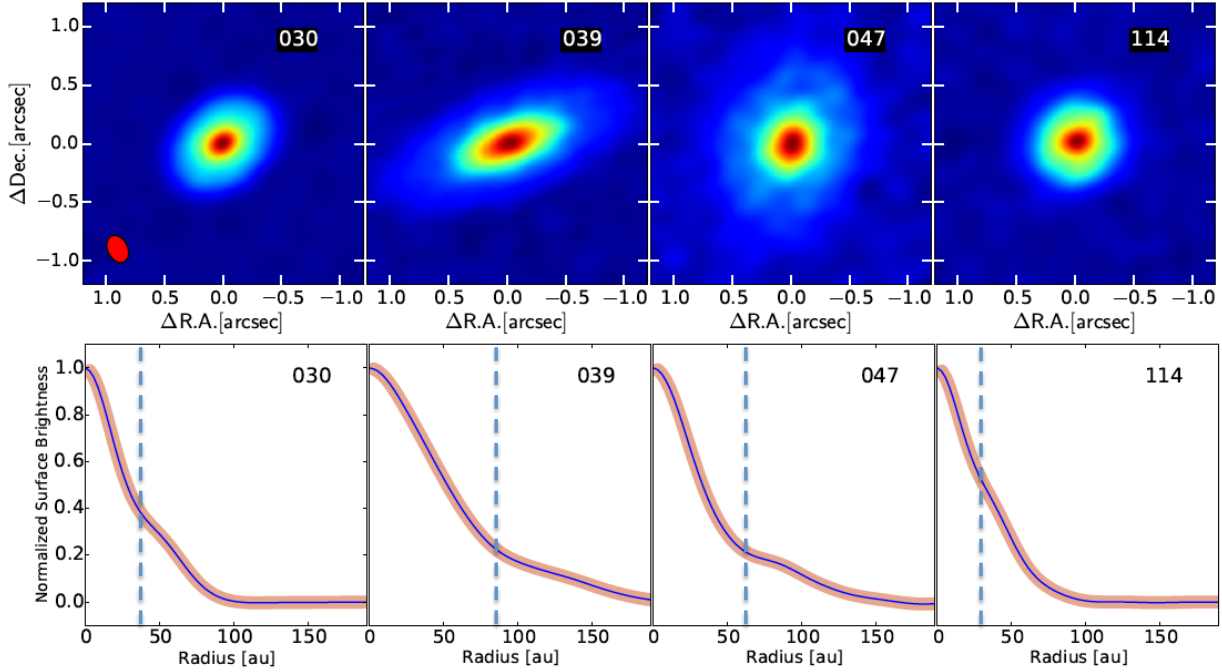}
\caption{ \small Images (top) and radial profiles (bottom) of the  ODISEA objects with hints of substructures.  
The vertical lines show changes in slope indicating possible substructures. 
}
\label{fig:possible_structures}
\end{figure*}

To search for additional, more subtle substructures,  we use the coordinates, position angles and FWHM values of the major and minor axes listed in Table~\ref{table:results}  to deproject the images and plot the deprojected radial profiles. Figures~\ref{fig:smooth1} shows the images with uniform weightings and deprojected radial profiles for all targets with fluxes $\gtrsim$15 mJy and errors in PA  $<$ 90 deg that are consistent with smooth disc structures. 
In Figure~\ref{fig:possible_structures}, we show the images and profiles of 4 sources where the radial profiles have breaks in slope, which we interpret as possible hints for  sub-structures (e.g., unresolved gaps).
They are objects number  30, 39, 47 and 114, which are best known as ROXR1 16, DoAr 25, ISO-Oph 54, and WSB 60. 
Using the same approach, DoAr 25 was already identified as a disc with possible substructure by Cox et al. (2017).
These four objects are among the brightest targets in the sample ($\gtrsim$100 mJy); therefore, if these substructures turn out to be real, 
it would imply that  $\sim$50$\%$ (11/21) of the sources brighter than $\gtrsim$50 mJy show some kind of substructure.

\section{Discussion}

\subsection{Disc dust masses and sizes}\label{dust_masses}

\begin{figure*}
\includegraphics[width=8.5cm, trim = 10mm 00mm 00mm 00mm, clip]{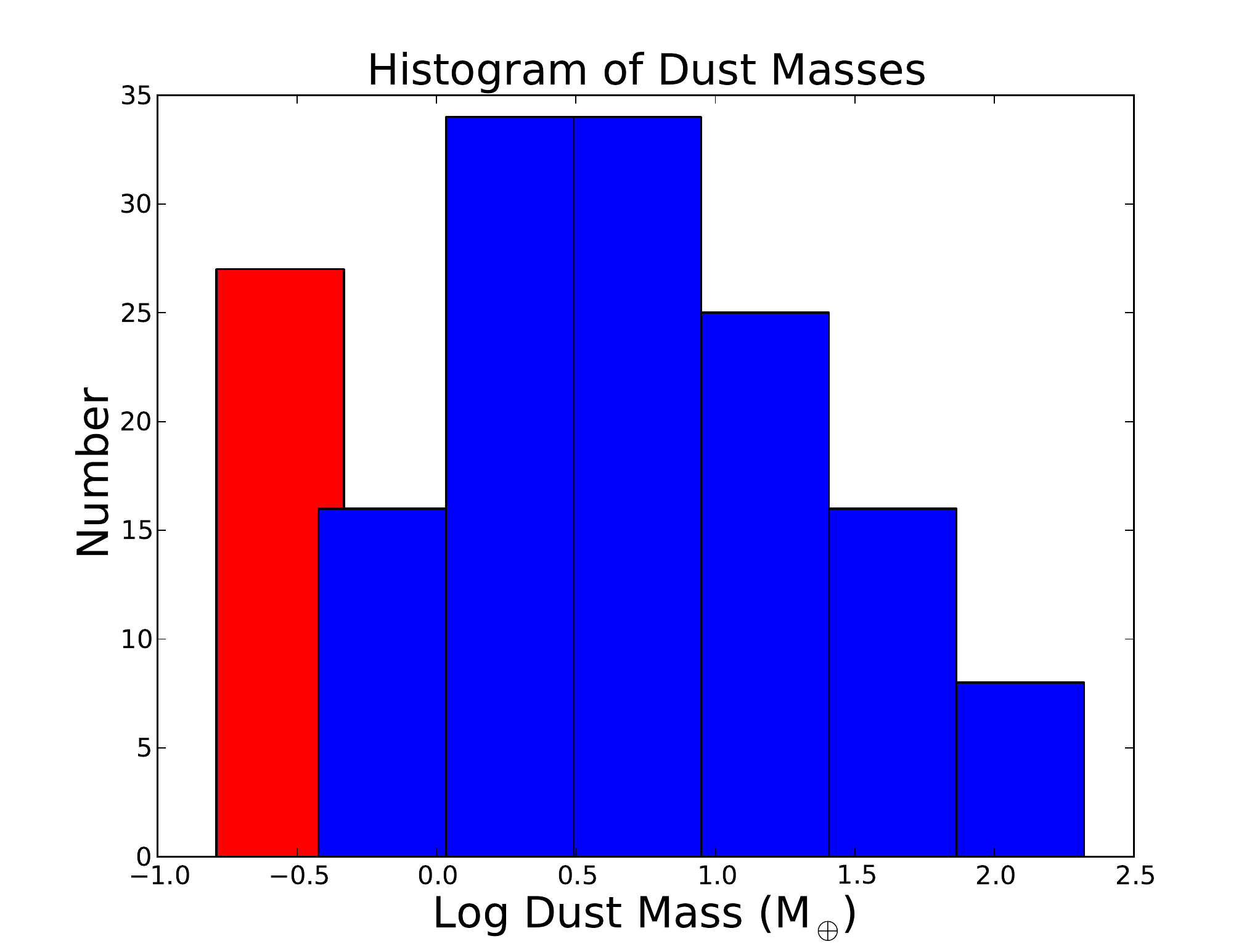}
\includegraphics[width=8.5cm, trim = 10mm 00mm 00mm 00mm, clip]{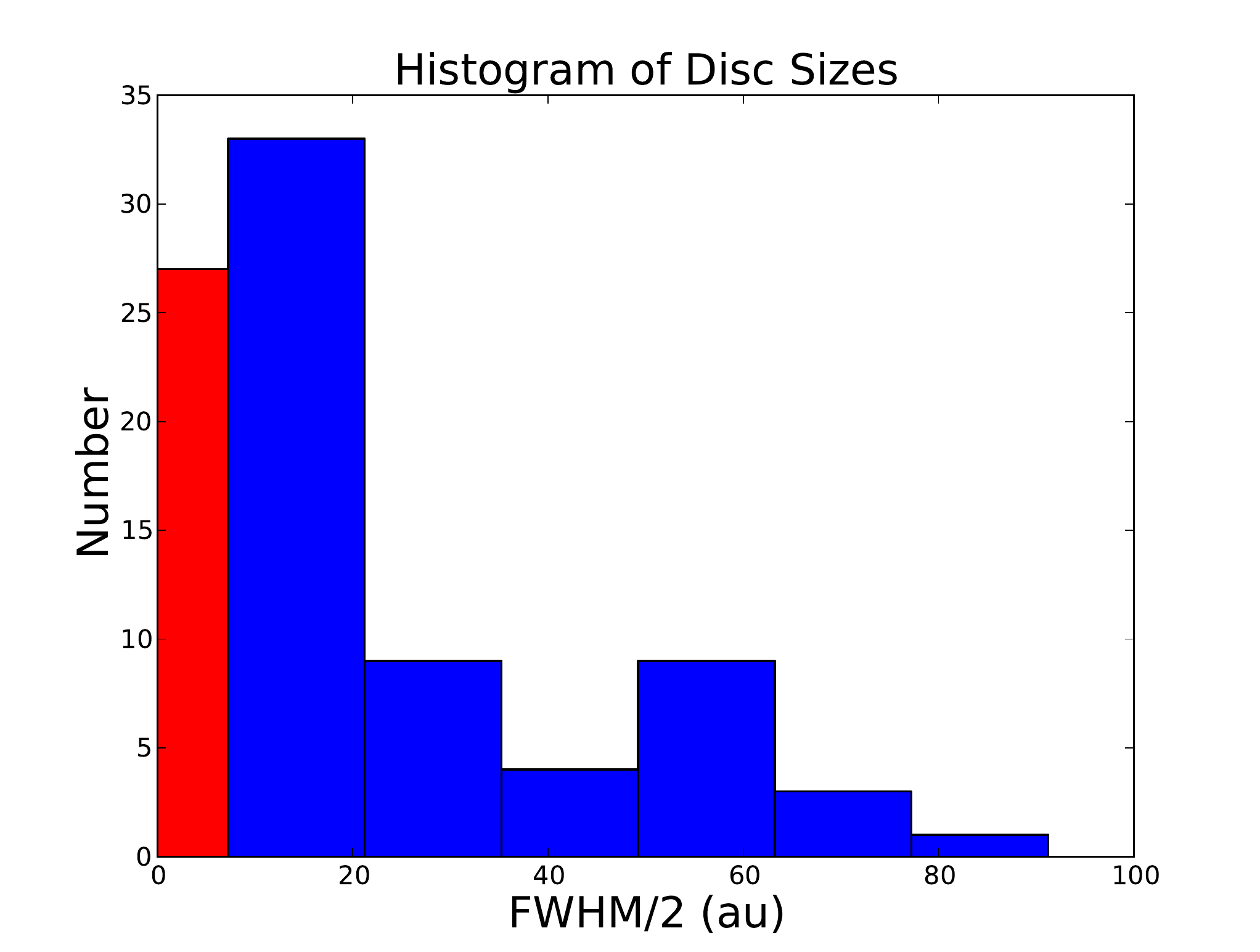}
\caption{ \small 
{\bf{Left Panel:}} histogram of dust masses derived for our 133 detected discs.
Mass upper limits (3-$\sigma$) for the 27 non-detections are shown in red (first bar from the left). 
{\bf{Right Panel:}} histogram of dust disc radii, defined as 0.5 $\times$ FWHM of the major axis, and  adopting  a distance of 140 pc.
A radius upper limit of 14 au has been assigned to unresolved sources (first bar from the left). 
}
\label{fig:disc-masses}
\end{figure*}

Since protoplanetary discs typically become optically thin at millimeter wavelengths outside the first few au, 
most dust particles in the disc contribute to the observed 1.3 mm fluxes; therefore, 
these values can be used to estimate the amount of dust present, as follows: 

\begin{equation}
M_{dust}  = \frac{F_{\nu} d^2}{\kappa_{\nu} B_{\nu} (T_{\rm dust})}
\end{equation}

Where $F_{\nu}$ is the flux, $B_{\nu}$ is the Plank function and $\kappa_\nu$ is the dust opacity. 
Adopting the prescription for   $\kappa_\nu$ at millimeter wavelengths from Beckwith et al. (1990),

\begin{equation}
\kappa_{\nu}  = 0.1 \left( \frac{\nu}{10^{12} \mathrm{Hz}} \right)^\beta \mathrm{cm^2  g^{-1}}
\end{equation}

we obtain $\kappa_{1.3mm}$ = 0.023 cm$^2$  g$^{-1}$ for $\beta$ = 1.0. If we further adopt $d$ = 140 pc  based on recent VLBA and Gaia results (Ortiz-Leon et al. 2017; Canovas et al., in prep.) and $T_{\rm dust}$ = 20 K, which is the median disc temperature in Taurus calculated by Andrews $\&$ Williams (2005), Equation 2 then becomes:

\begin{equation}
M_{dust}  = 0.58 \times \frac{F_{1.3mm}}{\mathrm{mJy}}  M_{\oplus}
\end{equation}

The linear relation between flux and dust mass is supported by the fact that 
the Planck function is close to the Rayleigh-Jeans regime at millimeter wavelengths, $B_{\nu} \sim 2 \nu^2  \kappa T /c^2$, and the emission
is only linearly (rather than exponentially) dependent on the dust temperature.
While the average disc temperature might be lower than 20 K in the discs  around brown dwarfs (van der Plas et al. 2016),  as shown by Tazzari et al. (2017),  the disc temperature does not depend strongly on stellar properties in the 0.1 to 2 M$_\odot$ stellar mass range. 
However,  the optically thin assumption breaks down in the  dense inner regions (Andrews $\&$ Williams, 2007), and depending on the size of the disc and its surface density profile,
the fraction of optically thick material might be significant even at 1.3 mm, resulting in \emph{underestimated} disc masses.
On the other hand, very compact discs around bright stars might have higher  average temperatures than 20 K.  If the temperatures are higher than the adopted 20 K value,  disc masses would then be \emph{overestimated}. 
Obtaining accurate masses for small and very dense discs might require high-resolution observations at longer wavelengths and detailed radiative transfer modeling. 
Fortunately, this is within the reach of ALMA capabilities,  which can  obtain 3.0 mm images at 0.04$''$ resolution (6 au at  140 pc). 
Disc masses from radiative transfer modeling of our sources will be presented by  Perez et al., (in preparation), but  
meanwhile, we estimate disc masses from Equation~3, which are shown in Figure~\ref{fig:disc-masses} (left panel).
Using Equation~3, we also find that the 0.16 mJy detection in the stacking analysis (Section~\ref{stacking}) corresponds to an average dust mass of just 0.09 M$_{\oplus}$.

Besides the mass, the size is another fundamental property of a protoplanetary disc.  Since protoplanetary discs do not have sharp edges, defining a size is not trivial. 
The large-scale radial structures of protoplanetary discs are often described by a characteristic radius and an exponential tapper in the outer disc (Hughes et al. 2008; Andrews et al. 2010;  Cieza et al. 2018). 
Characteristic radii for all the spatially resolved sources will also be derived from radiative transfer modeling, but for now we simply show  in Figure~\ref{fig:disc-masses} (right panel) the distribution disc sizes derived in Section~\ref{continuum}. Here we define the radius as FWHM/2 and show the values in au for the adopted distance of 140 pc. 
We emphasize that the disc sizes defined this way are not equivalent to disc outer radii, which are typically very difficult to quantify. 
Depending on the signal to noise of the data, interferometers like ALMA are able to measure the size of a disc that is slightly smaller than the beam by deconvolving  the beam from the signal. Therefore, relatively bright sources (e.g., S/N $\gtrsim$ 30) that are consistent with point sources are likely to be significantly smaller than the beam. For simplicity, we 
conservatively set the sizes of all the unresolved sources to a FWHM of 0.2$''$, corresponding to a radius of $\sim$14 au.

\subsection{Comparison to other regions}

\begin{figure*}
\includegraphics[width=15cm, trim = 0mm 0mm 0mm 0mm, clip]{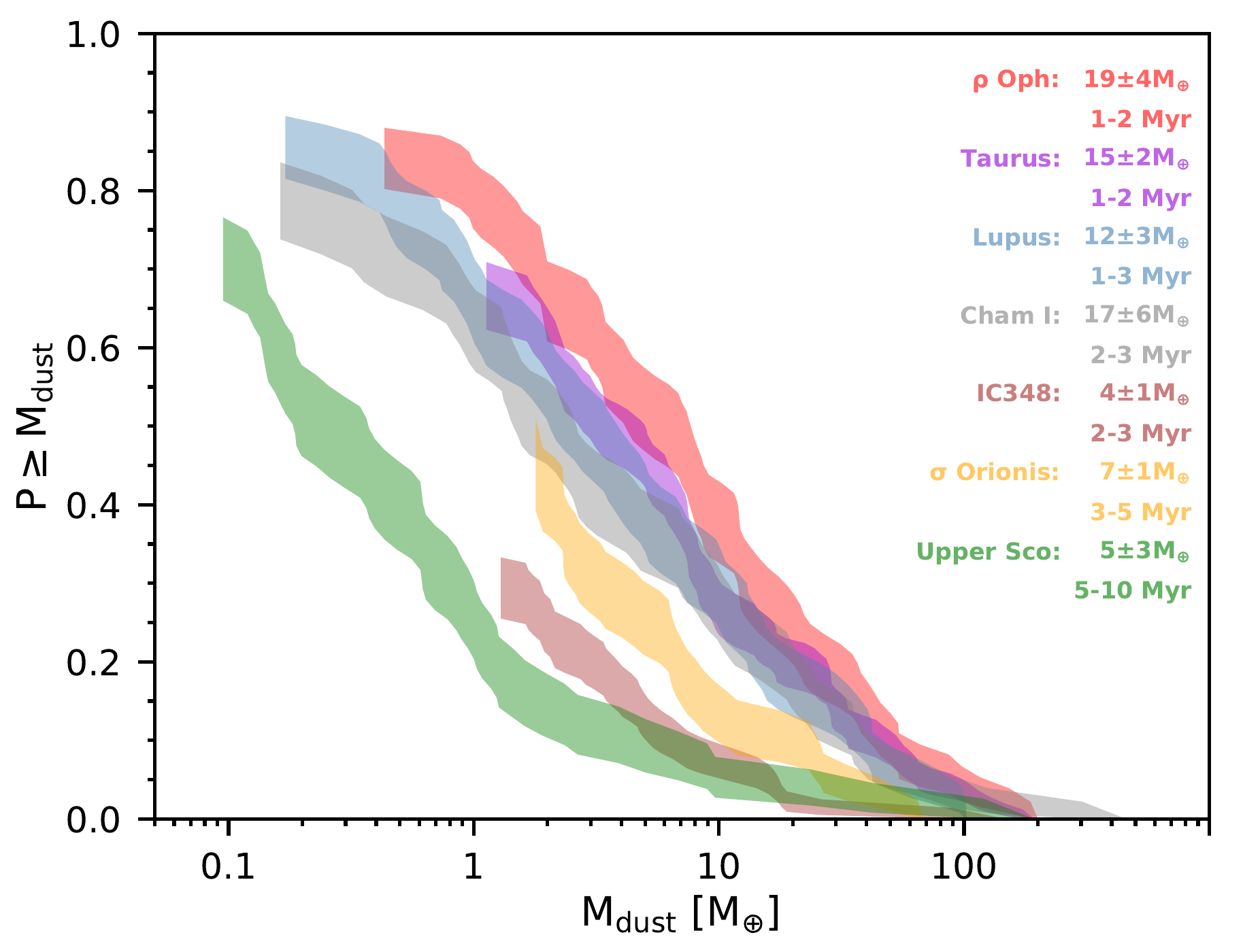}
\caption{ \small 
The cumulative dust mass distribution in our Ophiuchus sample compared to those of other young regions. 
The line widths indicate 1-$\sigma$ confidence intervals. 
Average dust masses and ages are listed for reference. 
Ophiuchus presents a disc mass distribution that is very similar to those seen Taurus and Lupus, and more massive discs than older regions. 
However, the distributions shown are not corrected by sample selection biasses or the dependence on stellar mass and thus should be interpreted with caution.  
}
\label{fig:comparison}
\end{figure*}

As discussed in Section~1, ALMA has already surveyed many of the nearby star-forming regions and  young clusters. 
This allows us to compare the distribution of dust masses seen in Ophiuchus to those in regions of different ages.  
In Figure~\ref{fig:comparison}, we show the cumulative distribution of the dust masses seen in Ophiuchus calculated 
using the Kaplan-Meier estimator in the ASURV package (Lavalley et al. 1992) to include upper limits, as described by Ansdell et al. (2016).
For comparison, we perform the same analysis for objects in Lupus (Ansdell et al.  2016), Upper Scorpius (Barenfeld  et al. 2016),   Chamaeleon I (Pascucci et al.  2016), 
 $\sigma$ Ori (Ansdell et al. 2017),  and IC~348 (Ruiz-Rodriguez et al. 2018). 
We also include the pre-ALMA study of Taurus (Andrews et al. 2013).  The figure shows the ages for each region adopted in each one of the (sub)millimeter surveys, but note 
that these ages were not necessarily calculated in fully consistent ways. 

We find that the dust mass distribution in our Ophiuchus sample (which is not yet complete) is very similar to those of other young regions (1-2 Myr) as Taurus and Lupus.  However, significant evolution is seen toward older ages, as shown by recent studies (Ansdell et al.  2017; Ruiz-Rodriguez et al. 2018). 
Given the strong dependence of disc masses on stellar mass (Andrews et al. 2013), properly comparing the distribution of dust masses requires controlling the disc sample for stellar mass. 
For instance, the dependence on stellar mass explains why the discs in IC~348, a cluster dominated by very-low-mass stars, appear much fainter than the discs in Cham II,  even though both clusters have a similar age.
Their dust mass distributions are much more  similar when accounting for the stellar mass dependence (Ruiz-Rodriguez et al. 2018).  
Spectral types can be used as a proxy for stellar mass but this becomes problematic when the range of stellar masses and ages increase because stars of a given stellar mass (specially solar-mass and higher mass stars) do change spectral types during their pre-main-sequence evolution.  
Given these caveats,  a more detailed comparison  of the dust distribution seen in Ophiuchus to other regions will be presented by Ruiz-Rodriguez et al. (in preparation) using individually derived stellar masses and the full Ophiuchus sample.  Based on the trends seen in Figure~\ref{fig:colormag-detec} (lower disc masses toward lower K-band fluxes and bluer [K]-[24] colors), we anticipate that the dust mass distribution of Ophiuchus discs should move toward lower dust masses when Sample B is included. 

\subsection{Implications for planet formation: challenges and future directions}\label{implications}

One of the main motivations of disc demographic studies like ODISEA is to inform planet-formation models and constrain the planet-formation potential of protoplanetary discs in
nearby molecular clouds, which are very challenging tasks.
While ALMA has unprecedented capabilities, it still provides limited information on the properties of protoplanetary discs and planet formation processes.  
The formation of rocky planets involves the growth of solid bodies over more than 12 orders of magnitude, from $\sim$10$^{-6}$  to $>$10$^{6}$ m scales (Lissauer, 1993). 
However, ALMA is mostly sensitive  to the emission of dust particles that are $\sim$10$^{-3}$ m in size. 
Given the size distributions that are usually adopted for protoplanetary discs, with n(a)  $\propto a^{-3.5}$  (Mathis, Rumpl $\&$ Nordsieck 1977) and a$_{\rm min}$  and a$_{\rm max}$ in the 
$\mu$m and mm regimes, respectively,  most of the mass is contributed by the largest bodies in the distribution, and the total dust mass scales as $a_{\rm max}^{1/2}$.
Therefore, (sub)millimeter fluxes constrain the mass of solids with sizes up to a few times the wavelength of the observations, but provide little information on the presence 
of larger bodies, beyond cm scales. 
This situation is complicated by the fact that radial drifting is a strong function of particle size  (Birnstiel et al. 2012) and thus the grain-size distribution is a strong function of radius. 
Furthermore, snow lines produce radial discontinuities in the distribution of particle sizes (Banzatti et al.  2015). 
Similarly, gas giant planets are mostly made of H$_2$, which ALMA can not directly observe.  Instead, the gas content of protoplanetary discs is usually probed with other gas tracers 
like CO isotopologues, which are difficult to interpret (Miotello et al. 2017). 

Very large objects might become detectable by ALMA, although indirectly, when they become massive enough to dynamically clear gaps in the disc. The minimum gap-opening mass depends on the  viscosity  and scale-height of the disc  (Duffell $\&$ MacFadyen, 2013) but in any case requires fully-formed planets, billions of times larger (in diameter) than the largest grains that are directly detectable by ALMA. 
Gaps consistent with being dynamically carved by fully-formed  planets have been imaged by ALMA in discs with estimated ages ranging from $\lesssim$ 1 Myr (HL Tau and Elias 2-24; ALMA Partnership et al. 2015,    Cieza et al. 2017) to $\sim$10 Myr  (TW Hydra; Andrews et al. 2016).  
However,  the origin of these gaps still remain to be established and several alternative explanations have been proposed, including the snow-lines of different species (Zhang et al.  2015; Okuzumi et al.  2016), magneto-hydrodynamic effects (Ruge et al.  2016, Flock et al.  2017), and viscous ring-instabilities (Dullemond $\&$ Penzlin, 2018).

Despite all the caveats discussed above, large ALMA surveys of protoplanetary discs still provide critical information on planet formation. 
The (sub)millimeter fluxes constrain the amount of raw solid material available for the formation of planets. In particular, the dust masses derived in Section~\ref{dust_masses} represent a lower limit to the amount of  solids present in a given system.  Also,  relative dust masses are expected to be less uncertain than the absolute values, allowing the study of meaningful  correlations between dust masses and stellar properties such as mass (Andrews et al. 2013),  age (Ruiz-Rodriguez et al. 2018),  multiplicity (Cox et al. 2017) and environment (Mann et al. 2014).
As mentioned in Section~\ref{introduction}, such correlations will be investigated in future ODISEA papers. 

Based on the results from Section~\ref{dust_masses}, we find that all detected sources (120 targets with $F_{1.3mm} \gtrsim$ 1 mJy, or 82$\%$ of the sample) have enough solids (in the form of dust) to form rocky planets ($M_{dust}$ $\gtrsim$ 0.5 M$_{\oplus}$). 
In contrast, only $\sim$50 targets  (1/3 of the sample) have enough dust to form  a critical-mass rocky core massive enough to start runaway  gas accretion ($\sim$10 M$_{\oplus}$, Lissauer 1993; Pollack et al. 1996) and trigger the formation of a gas giant. 
While these connections between disc properties and planet formation potential are just first-order approximations, they are in broad agreement with the relative incidence of extra-solar planets  (Cassan et al. 2012; Howard, 2013; Burke et al. 2015; Shvartzvald et al. 2016).  

Other disc properties such as radius  are less uncertain than absolute discs masses.  The distribution of disc sizes shown in Figure~\ref{fig:disc-masses} (right panel) suggests that planetary systems with compact architectures (e.g., most planets within 20 au) should be much more common than systems with planets in broad ($a$ $>$ 20 au) orbits, also in agreement with current constrains on extra-solar planets
at these wide separations (Bowler 2016; Vigan et al. 2017).  
Similarly, if gaps and/or cavities like those discussed in Section~\ref{structures} are in fact due to planets (e.g., Keppler et al. 2018), those features might be used to inform the incidence of different types of planets at different radii. 

As disc studies are completed with ALMA and the searches for extrasolar planets continue to expand the parameter space (in terms of planet mass and semi-major axis), it will become easier to connect both fields. However, given the limitations of the disc observations mentioned above, which are unlikely to be solved in the foreseeable future,  numerical modeling will remain the main way to investigate how  the small dust particles that are observable by radio telescopes can grow into fully-formed planets. 
In this context, disc demographic studies such as ODISEA can provide  valuable input, in the form of basic disc parameters and their dependence on stellar properties,  to planet population synthesis models (Alibert et al., 2005; Mordasini et al., 2009; Ronco et al. 2015),  which can then be used to connect disc populations to the populations of planets we see in the Galaxy. 

\section{Summary}

As part of the Ophiuchus disc Survey Employing ALMA (ODISEA) project, we have observed 147 discs at 0.2$''$ (28 au) resolution in 1.3 mm continuum and CO  isotopologues.  In this first paper, we describe the scope of the survey, which aims to study the entire population of $\sim$300 protoplanetary discs identified by the ``Cores to Discs"  \emph{Spitzer} Legacy Project in the Ophiuchus Molecular Cloud, and present the initial continuum data.  Our main results are the following: \\

\noindent 1) We detect 120 of our 147  targets in 1.3 mm  continuum, for an overall detection rate of  82$\%$.  Among these detections, we find 11 binary systems and a triple system, for a total of 133 individual discs detected. Out of these 133 detected discs,  53 were spatially resolved and we measure fluxes,   sizes,  and position angles.  The other 80 detected objects  remain unresolved and we can only measure flux and set an upper limit to their size (approximately corresponding to the resolution of our observations). \\

\noindent 2)  27 of our targets remain undetected, with 4-$\sigma$ upper limits of $\sim$1.0 mJy, but a stacking of the non-detections show a 4-$\sigma$ detection of 0.16 mJy, suggesting a typical dust mass of just  $\sim$0.1 M$_{\oplus}$ for these objects. \\

\noindent 3) Among our sample, we find 8 sources with clear substructures. Six of them have inner dust opacity cavities and two of them show narrow gaps/rings (one object, WSV~82 show both an inner cavity and a gap/ring structure). Another object, Elias 2-27, shows two spiral ams. Four additional sources show hints of substructures based on their deprojected radial profiles.   Most of these features are seen among the brightest sources in the sample.  If they are all confirmed,  it would imply that $\sim$50$\%$ (11/21) of the sources brighter than  $\sim$50 mJy show some type of substructure. \\

\noindent 4)  We performed a preliminary comparison of the dust mass function in Ophiuchus to those of other regions.
We find that the dust mass distribution in Ophiuchus is very similar to those of other young regions (1-2 Myr) as Taurus and Lupus.  However, significant evolution is seen toward older ages,
as already shown by previous results. \\

\noindent 5) A simple conversion between flux and dust mass (adopting standard assumptions for dust opacities and temperatures) indicate that all sources detected at 1.3 mm  have enough solid mass to form one or more rocky planets.  In contrast, only $\sim$50 discs ($\sim$1/3 of the sample) have enough mass in the form of dust to form the canonical 10 M$_{\oplus}$ core needed to trigger runaway gas accretion and the formation of gas giant planets. In this context,  the main uncertainty is the total mass of solids already incorporated to large bodies (e.g., cm to km scales) that are not detectable by ALMA.   \\

\noindent 6) The distribution in disc sizes in our sample is heavily weighted towards compact discs.  Most discs  have radii  $<$ 15 au, while only 22 discs ($\sim$15$\%$ of the targets) have radii $>$ 30 au.  
The discs that remain unresolved in our sample would benefit from higher-resolution data at longer wavelengths to better constrain both their sizes and masses.  \\

The detailed study of disc properties as a function of the mass and age  of the host stars, the effects of (sub)stellar companions on disc properties, and the gas content in the discs based on the $^{12}$CO, $^{13}$CO, and C$^{18}$O observations will be presented in future papers of this series, along with radiative transfer modeling of resolved sources.

\section*{Acknowledgments}
We thank the anonymous referee for his/hers comments, which we believe have helped us to improve the manuscript significantly.
This paper makes use of the following ALMA data: ADS/JAO.ALMA \#2016.1.00545.S.  ALMA is a partnership of ESO (representing its member states), NSF (USA) and NINS (Japan), together with NRC (Canada), NSC and ASIAA (Taiwan), and KASI (Republic of Korea), in cooperation with the Republic of Chile. 
The Joint ALMA Observatory is operated by ESO, AUI/NRAO and NAOJ.
The National Radio Astronomy Observatory is a facility of the National Science Foundation operated under cooperative agreement by Associated Universities, Inc.
L.A.C., S.C., G.H.M.B,  A.O, and A.Z  were supported by CONICYT-FONDECYT grant numbers 1171246,  1171624, 3170204, 1151512, and 3170657.
D.A.R. acknowledges support from NASA Exoplanets program grant NNX16AB43G. 
KPR acknowledges CONICYT PAI Concurso Nacional de Inserci\'on en la Academia, Convocatoria 2016 Folio PAI79160052.
L.A.C., S.C., S.P. and A.Z. acknowledge support from the  Millennium Nucleus ``Protoplanetary Discs in ALMA Early Science'', grant number RC130007.
A.B., J.O., M.R.S. acknowledge support from the  Millennium Nucleus for Planet Formation. 
G.H.M.B, also received funding from the European Research Council (ERC, grant agreement No. 757957).
GvdP acknowledges funding from ANR of France under contract number ANR-16-CE31-0013 (Planet-Forming-Disks)

\bsp

\label{lastpage}


\begin{thebibliography}{99}
\bibitem[Alibert et al.(2005)]{2005A&A...434..343A} Alibert, Y., Mordasini, C., Benz, W., \& Winisdoerffer, C.\ 2005, A$\&$A, 434, 343 
\bibitem[Allers et al.(2006)]{2006ApJ...644..364A} Allers, K.~N., Kessler-Silacci, J.~E., Cieza, L.~A., \& Jaffe, D.~T.\ 2006, ApJ, 644, 364 
\bibitem[ALMA Partnership et al.(2015)]{2015ApJ...808L...3A} ALMA Partnership, Brogan, C.~L., P{\'e}rez, L.~M., et al.\ 2015, ApJL, 808, L3 
\bibitem[Andrews \& Williams(2005)]{2005ApJ...631.1134A} Andrews, S.~M., \& Williams, J.~P.\ 2005, ApJ, 631, 1134 
\bibitem[Andrews \& Williams(2007)]{2007ApJ...671.1800A} Andrews, S.~M., \& Williams, J.~P.\ 2007, ApJ, 671, 1800 
\bibitem[Andrews et al.(2009)]{2009ApJ...700.1502A} Andrews, S.~M., Wilner, D.~J., Hughes, A.~M., Qi, C., \& Dullemond, C.~P.\ 2009, ApJ, 700, 1502 
\bibitem[Andrews et al.(2010)]{2010ApJ...723.1241A} Andrews, S.~M., Wilner, D.~J., Hughes, A.~M., Qi, C., \& Dullemond, C.~P.\ 2010, ApJ, 723, 1241 
\bibitem[Andrews et al.(2013)]{2013ApJ...771..129A} Andrews, S.~M., Rosenfeld, K.~A., Kraus, A.~L., \& Wilner, D.~J.\ 2013, ApJ, 771, 129 
\bibitem[Andrews et al.(2016)]{2016ApJ...820L..40A} Andrews, S.~M., Wilner, D.~J., Zhu, Z., et al.\ 2016, ApJL, 820, L40 
\bibitem[Ansdell et al.(2016)]{2016ApJ...828...46A} Ansdell, M., Williams, J.~P., van der Marel, N., et al.\ 2016, ApJ, 828, 46 
\bibitem[Ansdell et al.(2017)]{2017AJ....153..240A} Ansdell, M., Williams, J.~P., Manara, C.~F., et al.\ 2017, AJ, 153, 240 
\bibitem[Ansdell et al.(2018)]{2018ApJ...859...21A} Ansdell, M., Williams, J.~P., Trapman, L., et al.\ 2018, ApJ, 859, 21 
\bibitem[Banzatti et al.(2015)]{2015ApJ...815L..15B} Banzatti, A., Pinilla, P., Ricci, L., et al.\ 2015, ApJL, 815, L15 
\bibitem[Barenfeld et al.(2016)]{2016ApJ...827..142B} Barenfeld, S.~A., Carpenter, J.~M., Ricci, L., \& Isella, A.\ 2016, ApJ, 827, 142 
\bibitem[Bate(2018)]{2018MNRAS.475.5618B} Bate, M.~R.\ 2018, MNRAS, 475, 5618 
\bibitem[Beckwith et al.(1990)]{1990AJ.....99..924B} Beckwith, S.~V.~W., Sargent, A.~I., Chini, R.~S., \& Guesten, R.\ 1990, AJ, 99, 924 
\bibitem[Birnstiel et al.(2012)]{2012A&A...539A.148B} Birnstiel, T., Klahr, H., \& Ercolano, B.\ 2012, A\&A, 539, A148 
\bibitem[Bowler(2016)]{2016PASP..128j2001B} Bowler, B.~P.\ 2016, PASP, 128, 102001 
\bibitem[Burke et al.(2015)]{2015ApJ...809....8B} Burke, C.~J., Christiansen, J.~L., Mullally, F., et al.\ 2015, ApJ, 809, 8 
\bibitem[Casassus et al.(2013)]{2013Natur.493..191C} Casassus, S., van der Plas, G., M, S.~P., et al.\ 2013, Nature, 493, 191 
\bibitem[Cassan et al.(2012)]{2012Natur.481..167C} Cassan, A., Kubas, D., Beaulieu, J.-P., et al.\ 2012, Nature, 481, 167 
\bibitem[Chen et al.(1995)]{1995ApJ...445..377C} Chen, H., Myers, P.~C., Ladd, E.~F., \& Wood, D.~O.~S.\ 1995, ApJ, 445, 377 
\bibitem[Cieza et al.(2007)]{2007ApJ...667..308C} Cieza, L., Padgett, D.~L., Stapelfeldt, K.~R., et al.\ 2007, ApJ, 667, 308 
\bibitem[Cieza et al.(2009)]{2009ApJ...696L..84C} Cieza, L.~A., Padgett, D.~L., Allen, L.~E., et al.\ 2009, ApJL, 696, L84 
\bibitem[Cieza et al.(2010)]{2010ApJ...712..925C} Cieza, L.~A., Schreiber, M.~R., Romero, G.~A., et al.\ 2010, ApJ, 712, 925 
\bibitem[Cieza et al.(2012)]{2012ApJ...750..157C} Cieza, L.~A., Schreiber, M.~R., Romero, G.~A., et al.\ 2012a, ApJ, 750, 157 
\bibitem[Cieza et al.(2012)]{2012ApJ...752...75C} Cieza, L.~A., Mathews, G.~S., Williams, J.~P., et al.\ 2012b, ApJ, 752, 75 
\bibitem[Cieza et al.(2013)]{2013ApJ...762..100C} Cieza, L.~A., Olofsson, J., Harvey, P.~M., et al.\ 2013, ApJ, 762, 100 
\bibitem[Cieza et al.(2016)]{2016Natur.535..258C} Cieza, L.~A., Casassus, S., Tobin, J., et al.\ 2016, Nature, 535, 258 
\bibitem[Cieza et al.(2017)]{2017ApJ...851L..23C} Cieza, L.~A., Casassus, S., P{\'e}rez, S., et al.\ 2017, ApJL, 851, L23 
\bibitem[Cox et al.(2017)]{2017ApJ...851...83C} Cox, E.~G., Harris, R.~J., Looney, L.~W., et al.\ 2017, ApJ, 851, 83 
\bibitem[Currie \& Kenyon(2009)]{2009AJ....138..703C} Currie, T., \& Kenyon, S.~J.\ 2009, AJ, 138, 703 
\bibitem[Cutri et al.(2003)]{2003tmc..book.....C} Cutri, R.~M., Skrutskie, M.~F., van Dyk, S., et al.\ 2003, ``The IRSA 2MASS All-Sky Point Source Catalog, NASA/IPAC Infrared Science Archive.
\bibitem[Dipierro et al.(2018)]{2018MNRAS.475.5296D} Dipierro, G., Ricci, L., P{\'e}rez, L., et al.\ 2018, MNRAS, 475, 5296 
\bibitem[Duffell \& MacFadyen(2013)]{2013ApJ...769...41D} Duffell, P.~C., \& MacFadyen, A.~I.\ 2013, ApJ, 769, 41 
\bibitem[Dullemond \& Penzlin(2018)]{2018A&A...609A..50D} Dullemond, C.~P., \& Penzlin, A.~B.~T.\ 2018, A$\&$A, 609, A50 
\bibitem[Dunham et al.(2008)]{2008ApJS..179..249D} Dunham, M.~M., Crapsi, A., Evans, N.~J., II, et al.\ 2008, ApJS, 179, 249 
\bibitem[Erickson et al.(2011)]{2011AJ....142..140E} Erickson, K.~L., Wilking, B.~A., Meyer, M.~R., Robinson, J.~G., \& Stephenson, L.~N.\ 2011, AJ, 142, 140 
\bibitem[Evans et al.(2009)]{2009arXiv0901.1691E} Evans, N., Calvet, N., Cieza, L., et al.\ 2009a, arXiv:0901.1691 
\bibitem[Evans et al.(2009)]{2009ApJS..181..321E} Evans, N.~J., II, Dunham, M.~M., J{\o}rgensen, J.~K., et al.\ 2009b, ApJS, 181, 321 
\bibitem[Fern{\'a}ndez-L{\'o}pez et al.(2017)]{2017ApJ...845...10F} Fern{\'a}ndez-L{\'o}pez, M., Zapata, L.~A., \& Gabbasov, R.\ 2017, ApJ, 845, 10 
\bibitem[Flock et al.(2017)]{2017arXiv171006007F} Flock, M., Nelson, R.~P., Turner, N.~J., et al.\ 2017, arXiv:1710.06007 
\bibitem[Gaia Collaboration et al.(2018)]{2018arXiv180409365G} Gaia Collaboration, Brown, A.~G.~A., Vallenari, A., et al.\ 2018, arXiv:1804.09365 
\bibitem[Gaidos et al.(2016)]{2016MNRAS.457.2877G} Gaidos, E., Mann, A.~W., Kraus, A.~L., \& Ireland, M.\ 2016, MNRAS, 457, 2877 
\bibitem[G{\'a}sp{\'a}r \& Rieke(2014)]{2014ApJ...784...33G} G{\'a}sp{\'a}r, A., \& Rieke, G.~H.\ 2014, ApJ, 784, 33 
\bibitem[Greene et al.(1994)]{1994ApJ...434..614G} Greene, T.~P., Wilking, B.~A., Andre, P., Young, E.~T., \& Lada, C.~J.\ 1994, ApJ, 434, 614 
\bibitem[Hardy et al.(2015)]{2015A&A...583A..66H} Hardy, A., Caceres, C., Schreiber, M.~R., et al.\ 2015, A$\&$A, 583, A66 
\bibitem[Howard(2013)]{2013Sci...340..572H} Howard, A.~W.\ 2013, Science, 340, 572 
\bibitem[Hughes et al.(2008)]{2008ApJ...678.1119H} Hughes, A.~M., Wilner, D.~J., Qi, C., \& Hogerheijde, M.~R.\ 2008, ApJ, 678, 1119 
\bibitem[Keppler et al.(2018)]{2018arXiv180611568K} Keppler, M., Benisty, M., M{\"u}ller, A., et al.\ 2018, arXiv:1806.11568 
\bibitem[Kohn et al.(2016)]{2016ApJ...820....2K} Kohn, S.~A., Shkolnik, E.~L., Weinberger, A.~J., Carlberg, J.~K., \& Llama, J.\ 2016, ApJ, 820, 2 
\bibitem[Kraus et al.(2012)]{2012ApJ...745...19K} Kraus, A.~L., Ireland, M.~J., Hillenbrand, L.~A., \& Martinache, F.\ 2012, ApJ, 745, 19 
\bibitem[Lavalley et al.(1992)]{1992ASPC...25..245L} Lavalley, M., Isobe, T., \& Feigelson, E.\ 1992, Astronomical Data Analysis Software and Systems I, 25, 245 
\bibitem[Lissauer(1993)]{1993ARA&A..31..129L} Lissauer, J.~J.\ 1993, ARA$\&$A, 31, 129 
\bibitem[Loinard et al.(2008)]{2008ApJ...675L..29L} Loinard, L., Torres, R.~M., Mioduszewski, A.~J., \& Rodr{\'{\i}}guez, L.~F.\ 2008, ApJL, 675, L29 
\bibitem[Luhman \& Rieke(1999)]{1999ApJ...525..440L} Luhman, K.~L., \& Rieke, G.~H.\ 1999, ApJ, 525, 440 
\bibitem[Manara et al.(2015)]{2015A&A...579A..66M} Manara, C.~F., Testi, L., Natta, A., \& Alcal{\'a}, J.~M.\ 2015, A\&A, 579, A66 
\bibitem[Mann et al.(2014)]{2014ApJ...784...82M} Mann, R.~K., Di Francesco, J., Johnstone, D., et al.\ 2014, ApJ, 784, 82 
\bibitem[Mathis et al.(1977)]{1977ApJ...217..425M} Mathis, J.~S., Rumpl, W., \& Nordsieck, K.~H.\ 1977, ApJ, 217, 425 
\bibitem[McClure et al.(2010)]{2010ApJS..188...75M} McClure, M.~K., Furlan, E., Manoj, P., et al.\ 2010, ApJS, 188, 75 
\bibitem[McMullin et al.(2007)]{2007ASPC..376..127M} McMullin, J.~P., Waters, B., Schiebel, D., Young, W., \& Golap, K.\ 2007, Astronomical Data Analysis Software and Systems XVI, 376, 127 
\bibitem[Miotello et al.(2017)]{2017A&A...599A.113M} Miotello, A., van Dishoeck, E.~F., Williams, J.~P., et al.\ 2017, A$\&$A, 599, A113 
\bibitem[Mordasini et al.(2009)]{2009A&A...501.1161M} Mordasini, C., Alibert, Y., Benz, W., \& Naef, D.\ 2009, A$\&$A, 501, 1161 
\bibitem[Okuzumi et al.(2016)]{2016ApJ...821...82O} Okuzumi, S., Momose, M., Sirono, S.-i., Kobayashi, H., \& Tanaka, H.\ 2016, ApJ, 821, 82 
\bibitem[Ortiz-Le{\'o}n et al.(2017)]{2017ApJ...834..141O} Ortiz-Le{\'o}n, G.~N., Loinard, L., Kounkel, M.~A., et al.\ 2017, ApJ, 834, 141 
\bibitem[Owen(2016)]{2016PASA...33....5O} Owen, J.~E.\ 2016, PASA, 33, e005 
\bibitem[Pascucci et al.(2016)]{2016ApJ...831..125P} Pascucci, I., Testi, L., Herczeg, G.~J., et al.\ 2016, ApJ, 831, 125 
\bibitem[P{\'e}rez et al.(2016)]{2016Sci...353.1519P} P{\'e}rez, L.~M., Carpenter, J.~M., Andrews, S.~M., et al.\ 2016, Science, 353, 1519 
\bibitem[Pi{\'e}tu et al.(2014)]{2014A&A...564A..95P} Pi{\'e}tu, V., Guilloteau, S., Di Folco, E., Dutrey, A., \& Boehler, Y.\ 2014, A\&A, 564, A95 
\bibitem[Pollack et al.(1996)]{1996Icar..124...62P} Pollack, J.~B., Hubickyj, O., Bodenheimer, P., et al.\ 1996, Icarus, 124, 62 
\bibitem[Ratzka et al.(2005)]{2005A&A...437..611R} Ratzka, T., K{\"o}hler, R., \& Leinert, C.\ 2005, A$\&$A, 437, 611 
\bibitem[Rebollido et al.(2015)]{2015A&A...581A..30R} Rebollido, I., Mer{\'{\i}}n, B., Ribas, {\'A}., et al.\ 2015,  A$\&$A, 581, A30 
\bibitem[Ronco et al.(2015)]{2015A&A...584A..47R} Ronco, M.~P., de El{\'{\i}}a, G.~C., \& Guilera, O.~M.\ 2015, A\&A, 584, A47 
\bibitem[Ruge et al.(2016)]{2016A&A...590A..17R} Ruge, J.~P., Flock, M., Wolf, S., et al.\ 2016, A$\&$A, 590, A17 
\bibitem[Ru{\'{\i}}z-Rodr{\'{\i}}guez et al.(2016)]{2016MNRAS.463.3829R} Ru{\'{\i}}z-Rodr{\'{\i}}guez, D., Ireland, M., Cieza, L., \& Kraus, A.\ 2016, MNRAS, 463, 3829 
\bibitem[Ru{\'{\i}}z-Rodr{\'{\i}}guez et al.(2018)]{2018MNRAS.478.3674R} Ru{\'{\i}}z-Rodr{\'{\i}}guez, D., Cieza, L.~A., Williams, J.~P., et al.\ 2018, MNRAS, 478, 3674 
\bibitem[Shvartzvald et al.(2016)]{2016MNRAS.457.4089S} Shvartzvald, Y., Maoz, D., Udalski, A., et al.\ 2016, MNRAS, 457, 4089 
\bibitem[Tazzari et al.(2017)]{2017A&A...606A..88T} Tazzari, M., Testi, L., Natta, A., et al.\ 2017, A$\&$A, 606, A88 
\bibitem[Tripathi et al.(2017)]{2017ApJ...845...44T} Tripathi, A., Andrews, S.~M., Birnstiel, T., \& Wilner, D.~J.\ 2017, ApJ, 845, 44 
\bibitem[van der Marel et al.(2013)]{2013Sci...340.1199V} van der Marel, N., van Dishoeck, E.~F., Bruderer, S., et al.\ 2013, Science, 340, 1199 
\bibitem[van der Plas et al.(2016)]{2016ApJ...819..102V} van der Plas, G., M{\'e}nard, F., Ward-Duong, K., et al.\ 2016,  ApJ, 819, 102 
\bibitem[Vigan et al.(2017)]{2017A&A...603A...3V} Vigan, A., Bonavita, M., Biller, B., et al.\ 2017, A$\&$A, 603, A3 
\bibitem[Wahhaj et al.(2010)]{2010ApJ...724..835W} Wahhaj, Z., Cieza, L., Koerner, D.~W., et al.\ 2010, ApJ, 724, 835 
\bibitem[Wilking et al.(2005)]{2005AJ....130.1733W} Wilking, B.~A., Meyer, M.~R., Robinson, J.~G., \& Greene, T.~P.\ 2005, AJ, 130, 1733 
\bibitem[Williams \& Cieza(2011)]{2011ARA&A..49...67W} Williams, J.~P., \& Cieza, L.~A.\ 2011, ARA$\&$A, 49, 67 
\bibitem[Zhang et al.(2015)]{2015ApJ...806L...7Z} Zhang, K., Blake, G.~A., \& Bergin, E.~A.\ 2015, ApJL, 806, L7 
\end{thebibliography}
\end{document}